\documentclass[twocolumn,english,aps,prl,groupedaddress]{revtex4-2}
\usepackage[utf8]{inputenc}
\usepackage[T1]{fontenc}
\usepackage{xcolor}
\usepackage{verbatim}
\usepackage{amsmath}
\usepackage{amssymb}
\usepackage{graphicx}
\usepackage{textgreek}
\makeatletter

\DeclareRobustCommand{\greektext}{%
  \fontencoding{LGR}\selectfont\def\encodingdefault{LGR}}
\DeclareRobustCommand{\textgreek}[1]{\leavevmode{\greektext #1}}
\ProvideTextCommand{\~}{LGR}[1]{\char126#1}

\usepackage{hyperref}
\usepackage{placeins}
\renewcommand{\fnum@figure}{FIG.~\thefigure}
\makeatother

\usepackage{babel}
\begin{document}
\title{A unified perspective of high-harmonic generation in gases and solids }
\author{\textcolor{black}{A. Thorpe}\textsuperscript{\textcolor{black}{1}}}
\email{athor087@uottawa.ca}

\author{N. Boroumand\textsuperscript{1}}
\author{G. Bart\textsuperscript{1}}
\author{\textcolor{black}{L. Wang}\textsuperscript{\textcolor{black}{1}}}
\author{\textcolor{black}{G. Vampa}\textsuperscript{\textcolor{black}{1,2}}}
\author{\textcolor{black}{T. Brabec}\textsuperscript{\textcolor{black}{1}}}
\affiliation{\textsuperscript{1}Department of Physics, University of Ottawa, Ontario K1N 6N5, Canada}
\affiliation{\textsuperscript{2}National Research Council of Canada, Ottawa, Ontario K1A 0R6, Canada}
\date{\today}
\begin{abstract}
We present a quantum optical generalization of the quantum-matter Lewenstein model of high-harmonic generation 
(HHG) in gases that contains two channels corresponding to the inter- and intraband HHG in solids. Both channels can be presented as 
a semiclassical current multiplied by the vacuum field strength, resulting in a quantitative correction and a faster roll-off of the harmonic power with order than in previous theories. In gases, like in solids, intraband HHG dominates at low orders; the switchover harmonic corresponds to a specific photon energy, independent of pump wavelength. 
%The quantum optical Lewenstein model is a simple, scalable, and predictive tool for HHG design.
\end{abstract}
\maketitle

Theories of the interaction of intense electromagnetic radiation with matter
predate the widespread use of intense laser sources\citep{keldysh,faisal1973,reiss1973}.
One strong-field laser phenomenon, high-harmonic generation (HHG), was initially
observed in gases during the late 1980s and early 1990s \citep{mcpherson1987,ferray88,lhuillier1993}
and has since been extensively investigated for its applications,
such as providing a source of ultrafast coherent light and applications
in attosecond spectroscopy \citep{amini2019,corkum2007,krausz2009,scrinzi2005}.

The exploration of HHG has expanded to solids \citep{ghimire2011}, offering an extension of attosecond science to convenient tabletop media \citep{amini2019,vampa2014,ghimire2019,goulielmakis2022,park2022,yue2022}. HHG in solids had allowed 
for applications like probing various aspects of electronic band structure  
\citep{vampa2015a,luu2015,almalki2018,lanin2017}, imaging transition dipole moments \citep{wu2022} and Berry curvatures \citep{banks2017,luu2018,lv2021}, as well as, topological states of matter \citep{parks2023,bera2023} and phase transitions \citep{heide2022}. 

Recently, there has been growing interest in the quantum optical aspects of HHG. This has opened a 
number of novel applications, such as generation of Schr\"{o}dinger cat states
\citep{tsatrafyllis2017,lewenstein2021,riveradean2021}, photon bunching and antibunching
\citep{lemieux,gombkoto2021,riveradean2022,sloan2023,theidel2024,yi2025,stammer2024}, and generation 
of squeezed light \citep{theidel2024,yi2025,stammer2024} from HHG with classical and 
non-classical light sources, such as bright squeezed vacuum 
\citep{eventzur2023,rasputnyi2024,heimerl2024,eventzur2024,gorlach2023,boroumand2025}. 
It has also been shown that bound states may
influence the quantum nature of the emitted high-harmonic (HH) radiation \citep{varro2021,gorlach2020}.

HHG is typically elucidated in atomic gases according to Lewenstein's seminal work \citep{lewenstein1994}. This semiclassical approach—treating light classically and matter quantum mechanically—considers oscillations of a field-induced bound-continuum state dipole. It has 
an intuitive three-step explanation: an electron is ionized by a strong laser field, then propagates within the field-dressed 
continuum, and finally recombines with an ionic core, converting its excess energy into ultrafast pulses of light with 
odd frequency multiples of the driving laser's frequency \citep{corkum1993,lewenstein1994,corkum2007,schafer1993}. A similar 
mechanism has been studied in solids, known as the interband mechanism, where an electron is excited to a conduction band and 
a hole is promoted to a valence band, and they propagate before recombining to emit HH light \citep{vampa2014,parks2020,vampa2017,thorpe2023}.

In addition to the interband mechanism, there exists also an intraband mechanism in solids \citep{vampa2014,vampa2017}. This mechanism
describes the generation of harmonics as a response to charge motion within individual electronic bands.
The relation between the two mechanisms in atoms, analogous to interband and intraband HHG in solids, has not 
been explored yet.

Finally, while semiclassical models reproduce HHG spectra qualitatively very well, they are not quantitatively accurate. This limitation arises from their neglect of vacuum electric field fluctuations, which are essential for a full quantum description.

Here we develop a quantum optical generalization of the quantum-matter Lewenstein model which yields the following insights. 

First, the quantum optical strong field model inherently contains interband and intraband currents for both gases and solids. As such, 
it allows for a one-to-one connection between HHG in gases and solids. In atoms, the intraband mechanism is related to
Brunel HHG \cite{brunel90}.

Second, the quantum optical Lewenstein model yields semiclassical currents multiplied by the vacuum electric field. This is lost in 
semiclassical approaches, and allows for a close to quantitative description of HHG, as long as a Coulomb correction factor to 
ionization is included, and bound states can be neglected. 

Third, due to the frequency dependence of the vacuum electric field, the frequency scaling of HHG is also different to semiclassical 
theories. This results in a faster decrease of the harmonic power with order than in previous theories. 

Fourth, our analysis shows that, like in solids, the intraband current in atomic gases is important. It is the dominant source for 
low order HHG in atomic and molecular gases. The crossover harmonic between interband and intraband HHG relates to a specific
photon energy independent of the pump laser wavelength. 

Finally, the computational efficiency of the quantum optical Lewenstein model, in particular after using saddle point integration
\cite{krausz00}, allows for efficient integration into macroscopic phase matching models and thus, a scalable, close-to quantitative
description of HHG experiments in atomic and molecular gases. This makes it a valuable tool for the design and interpretation of
experiments.

To model the problem, we introduce the Hamiltonian
$\hat{H}=\hat{H}_{m}+\hat{H}_{f}+\hat{H}_{i}$, where $\hat{H}_{m}=\hat{\mathbf{p}}^{2}/(2m)+V(\mathbf{x})$
is the matter Hamiltonian, $\hat{H}_{f}=\sum_{{\kappa}}\hbar\omega_{{\kappa}}\hat{a}_{{\kappa}}^{\dagger}\hat{a}_{{\kappa}}$
is the field Hamiltonian, and $\hat{H}_{i}=\mathbf{r}\vert e\vert\hat{\mathbf{F}}(\mathbf{x})$ is
the interaction Hamiltonian. Here, $\hat{\mathbf{p}}$ represents
the momentum operator, $\mathbf{r}$ defines the relative coordinates
and $\mathbf{x}$ defines the centre of mass coordinates. Further,
$m$ and $\vert e\vert$ define the electronic mass and charge. The
creation and annihilation operators $\hat{a}_{{\kappa}}^{\dagger}$
and $\hat{a}_{{\kappa}}$ correspond to the electromagnetic
field mode $\kappa \equiv(\mathbf{k},s)$,
with wavevector $\mathbf{k}$ and $s=1 \text{ or } 2$ labeling two orthogonal unit vectors that define the polarization basis of the light field. The
quantized electric field and vector potential are given by,
\begin{subequations}
\begin{align}
\hat{\mathbf{F}}(\mathbf{x}) & =i\sum_{\kappa}\eta_{\kappa}\omega_{\kappa}\mathbf{e}_{\kappa}\left(\hat{a}_{\kappa}e^{i\mathbf{k\mathbf{x}}}-\hat{a}_{\kappa}^{\dagger}e^{-i\mathbf{kx}}\right),\\
\hat{\mathbf{A}}(\mathbf{x}) & =\sum_{\kappa}\eta_{\kappa}\mathbf{e}_{\kappa}\left(\hat{a}_{\kappa}e^{i\mathbf{k\mathbf{x}}}+\hat{a}_{\kappa}^{\dagger}e^{-i\mathbf{kx}}\right),
\end{align}
\end{subequations}
respectively, where $\mathbf{e}_{\kappa}$ is the unit vector along the polarization
direction, $\eta_{\kappa}=\sqrt{\hbar/(2\omega_{k}\varepsilon_{0}V)}$
is the vacuum vector field amplitude, $\omega_{k}$ is the mode frequency, $\varepsilon_{0}$ is the 
vacuum permittivity, and $V$ is the quantization volume.

We use the strong field and dipole approximations, considering a hydrogen-like atom with a ground state 
$\vert0\rangle$ and continuum states $\vert\mathbf{p}\rangle=1/(2\pi\hbar)^{3/2} 
\exp((i/\hbar)\mathbf{p}\mathbf{r})$ \citep{boroumand2022,keldysh,amini2019}. Doing so, we utilize the
Ansatz
\begin{align}
\vert\Psi(t)\rangle & =\vert0\rangle\otimes\vert\phi_{0}(t)\rangle+\int d^{3}p\vert\mathbf{p}\rangle\otimes\vert\phi_{\mathbf{p}}(t)\rangle,\label{eq:Ansatz}
\end{align}
where photon quantum states $\vert\phi_{0}(t)\rangle$
and $\vert\phi_{\mathbf{p}}(t)\rangle$, are associated with the ground
and continuum states respectively. By determining the evolution of
$\vert\phi_{0}(t)\rangle$ with the Schr\"{o}dinger equation, we can measure
HHG using the photon number expectation value $\langle\hat{n}\rangle(t)\approx\langle\phi_{0}(t)\vert\hat{n}\vert\phi_{0}(t)\rangle$. To measure HHG Eq. (\ref{eq:Ansatz}) can be
generalized to many atoms \citep{boroumand2025}, but for clarity we highlight the single atom case 
here.

We simplify solving the Schr\"{o}dinger equation by 
using the transformation $\vert\varphi_{0}(t)\rangle = \hat{D}_{\alpha}^{\dagger}\hat{U}_{v}^{\dagger}\hat{U}_{i}^{\dagger}\vert\phi_{0}(t)\rangle$,
with the interaction picture operator $\hat{U}_{i}=\exp\left(-i\hat{H}_{f}t/\hbar\right)$,
the velocity gauge operator $\hat{U}_{v}=\exp\left(i\vert e\vert\mathbf{r}\hat{\mathbf{A}}/\hbar\right)$,
and the displacement operator over modes of the driving laser field $\hat{D}_{\alpha}=\exp\left(\sum_{\kappa}\alpha_{\kappa}\hat{a}_{\kappa}^{\dagger}-\alpha_{\kappa}^{*}\hat{a}_{\kappa}\right)$, with $\alpha_\kappa$ being the corresponding single-mode complex coherent state amplitudes.
This separates the strong driving laser field and quantum optical
fields, excluding the fundamental harmonic in HHG analysis. The electric
field splits into $\tilde{\mathbf{F}}(t)=\boldsymbol{\mathcal{F}}(t)+\hat{\mathbf{F}}(t)$,
with $\boldsymbol{\mathcal{F}}(t)$ treated classically and $\hat{\mathbf{F}}(t)$
quantum optically. Similarly, $\tilde{\mathbf{A}}(t)=\boldsymbol{\mathcal{A}}(t)+\hat{\mathbf{A}}(t)$\textcolor{black}{{}
and Rabi frequency $\tilde{\Omega}(t)=\vert e\vert\mathbf{d}(\mathbf{p}_{t})\tilde{\mathbf{F}}(t)/\hbar=\Omega(t)+\hat{\Omega}(t)$
}with $\mathbf{p}_{t}=\mathbf{p}+\vert e\vert\mathcal{\boldsymbol{\mathcal{A}}}(t)$\textcolor{black}{{}
and transition dipole moment $\mathbf{d}(\mathbf{p})=\langle\mathbf{p}\vert\mathbf{r}\vert0\rangle$
\citep{Supplement,boroumand2022}.} Solving the Schr\"{o}dinger equation
leads to the Volterra integro-differential equation \citep{Supplement}:
\begin{align}
\partial_{t}\vert\varphi_{0}(t)\rangle & =-\int d^{3}p\hat{c}_{\mathbf{p}}^{\dagger}(t)\int_{t_{0}}^{t}dt^{\prime}\hat{c}_{\mathbf{p}}(t^{\prime})\vert\varphi_{0}(t^{\prime})\rangle,\label{eq:Volterra}
\end{align}
where, 
\begin{align}
\hat{c}_{\mathbf{p}}(t)= & \tilde{\Omega}(t) \hat{D}_{\sigma}^{\dagger}(t)\exp\left(iS(t)\right),\label{eq:cp}
\end{align}
with, 
\begin{align}
S(t)= & \frac{1}{\hbar}\int_{t_{0}}^{t}\left(\frac{\mathbf{p}_{\tau}^{2}}{2m}+E_{0}\right)d\tau,
\end{align}
the classical action. Further, the displacement
operator $\hat{D}_{\sigma}^{\dagger}(t)=\exp\left(-\sum_{\kappa}\sigma_{\kappa}\hat{a}_{\kappa}^{\dagger}-\sigma_{\kappa}^{*}\hat{a}_{\kappa}\right)$
consists of 
\begin{align}
\sigma_{\kappa}(t)= & \frac{\vert e\vert E_{v}}{\hbar}\overline{\sigma}_{\!{\kappa}}(t)e^{-i(\mathbf{k}\mathbf{x}-\omega_{k}t)}\\
\overline{\sigma}_{\kappa}(t)= & \frac{-i}{m\omega_{k}}\int_{t_{0}}^{t}\mathbf{e}_{{\kappa}}\mathbf{p}_{\mkern-1.5mu t^{\prime}}e^{-i\omega_{k}(t-t^{\prime})}dt',\label{eq:sigmabara}
\end{align}
where, $E_{v}=\omega_{k}\eta_{\kappa}$ is the vacuum electric field amplitude; $\sigma_{\kappa}(t)$ defines the dressing of continuum electrons with coherent state radiation modes.

While numerical methods \citep{stammer2024} like the variational iteration method (see
\citep{wazwaz2011}) can solve equations like Eq. (\ref{eq:Volterra}),
we use a simpler approximation for clarity. Following the Lewenstein
approach for ground state depletion, we assume the photon wavefunction
varies slowly and can be extracted from the integral \textcolor{black}{\citep{lewenstein1994}.} Additionally, assuming first-order
expansion of the operator expressions in Eq. (\ref{eq:Volterra}), we find that, 
\begin{align}
\vert\varphi_{0}(t)\rangle & \approx\exp\left(\sum_{\kappa}h_{\kappa}\hat{a}_{\kappa}^{\dagger}-h_{\kappa}^{*}\hat{a}_{\kappa}\right)\vert\varphi_{0}(t_{0})\rangle,\label{eq:phi0result}
\end{align}
where $h_{\kappa}$ is the coherent state parameter
that we will define in Eqs. (\ref{eq:hA}) and (\ref{eq:hB}) below. 
Assuming the initial state of the system is vacuum $\vert\varphi_{0}(t_{0})\rangle=\vert v\rangle$,
Eq. (\ref{eq:phi0result}) represents a coherent state. 

\textcolor{black}{To connect with semiclassical analysis of HHG, }the
coherent state parameter \textcolor{black}{$h_{\kappa}=h_{\kappa}^{\prime}+h_{\kappa}^{\prime\prime}$ involves} two channels\textcolor{black}{\label{hAB}
\begin{align}
h_{\kappa}^{\prime}= & \frac{\vert e\vert E_{v}}{\hbar}\mathbf{e}_{\kappa}\int_{-\infty}^{\infty}dte^{i(\omega_{k}t-\mathbf{k}\mathbf{x})}\mathbf{x}(t),\label{eq:hA}\\
h_{\kappa}^{\prime\prime}= & - \frac{\vert e\vert E_{v}}{\hbar}\int_{-\infty}^{\infty}dte^{i(\omega_{k}t-\mathbf{k}\mathbf{x})}\int d^{3}p\bar{\sigma}_{\kappa}(t)\left[\Gamma_{\mathbf{p}}(t)+\text{c.c.}\right],\label{eq:hB}
\end{align}
with,}
\begin{align}
\mathbf{x}(t) & =2\text{Re}\left[i\int d^{3}p\mathbf{d}^{*}(\mathbf{p}_{t})e^{-iS(t)}\int_{-\infty}^{t}dt^{\prime}\Omega(t^{\prime})e^{iS(t^{\prime})}\right],\label{eq:xGSLew}
\end{align}
\vspace{-12pt}
\begin{align}
\Gamma_{\mathbf{p}}(t)= & \Omega^{*}(t)e^{-iS(t)}\int_{-\infty}^{t}dt^{\prime}\Omega(t^{\prime})e^{iS(t^{\prime})}.\label{eq:Gamma}
\end{align}
Here, $\mathbf{x}(t)$ in Eq. (\ref{eq:xGSLew}) is Lewenstein's
time-dependent dipole which describes charge oscillations due to
polarization buildup between the atomic ground state and electronic
continuum states \citep{lewenstein1994}. Eq. (\ref{eq:Gamma}),
is the momentum-dependent complex ionization rate, related to the
full complex ionization rate $\gamma(t)=\int d^{3}p\Gamma_{\mathbf{p}}(t)$,
as defined in Lewenstein's work \citep{lewenstein1994}. Note that
in numerical calculations involving Eqs. (\ref{eq:hA})
and (\ref{eq:hB}) we use a filter to suppress electron returns longer than a half-cycle following ionization as the shorter trajectories are dominant in HHG in atomic gases (see Appendix C of supplement Ref. \citep{Supplement}).

The first mechanism, Eq. (\ref{eq:hA}), resembles
the Fourier transform of $\mathbf{x}(t)$ of Eq. (55) from Lewenstein's
work corresponding to polarization generated by the quantum interference between the laser-accelerated continuum electron and the atomic ground state. Eq. (\ref{eq:hA}) differs from Lewenstein's model through
the harmonic frequency factor $E_{v}\propto\omega_{k}^{1/2}$ due
to vacuum fluctuations, affecting polarization buildup and HHG.

The second mechanism, Eq. (\ref{eq:hB}), resembles the Fourier transform of the complex ionization rate 
$\gamma(t)=\int d^{3}p\Gamma_{\mathbf{p}}(t)$ of Eq. \textcolor{black}{(52)} of 
Lewenstein's work \citep{lewenstein1994}. Unlike the time dependent dipole $\mathbf{x}(t)$ found in the 
first channel, Eq. (\ref{eq:hA}), this source was not studied there as a source of HHG. Eq. (\ref{eq:hB}) allows HH emission through (i) ionization nonlinearities from $\Gamma_{\mathbf{p}}(t)$ and (ii) from dressing of the continuum electron as a coherent state, quantified by the parameter $\bar{\sigma}_{\kappa}(t)$ 
\citep{varro2021}. \textcolor{black}{The formalism and the equations introduced up to now have been derived in Ref. \citep{boroumand2025}. There, the focus was on HH emission resulting from Eq. (\ref{eq:hA}). Below we analyze HH emission from Eq. (\ref{eq:hB}) and we formally show the correspondence with intraband and interband currents.}
\begin{figure}[b]
\textcolor{black}{\vspace{-10pt}}\includegraphics{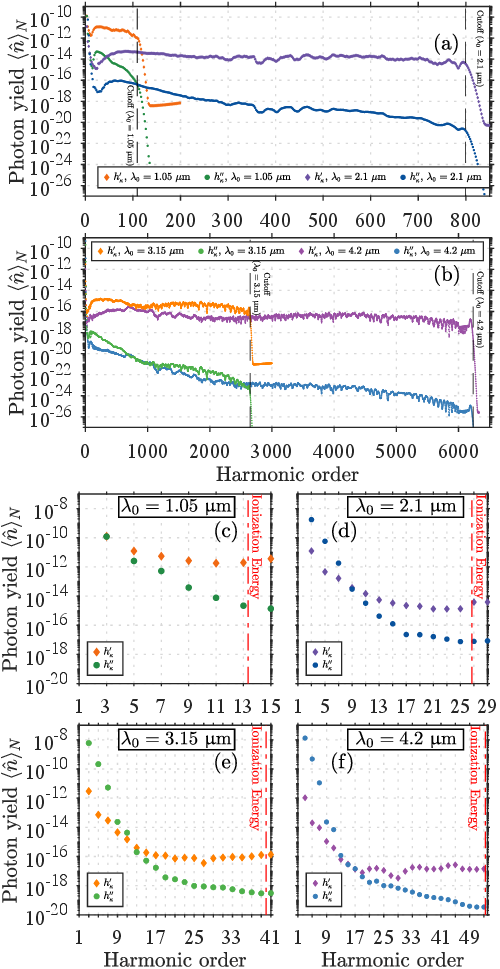}\caption{Harmonic spectra, calculated via Eq. (\ref{eq:yield}),
for wavelengths 1.05 \textgreek{m}m (a,c) and 2.1 \textgreek{m}m (a,d)
and 3.15 \textgreek{m}m (b,e) and 4.2 \textgreek{m}m (b,f) using a
Gaussian temporal pulse (5-cycle temporal FWHM) with intensity $3.5\times10^{14}\text{W}/\text{cm}^{2}$
in argon gas. Panels (a,b) show full spectra \textcolor{black}{up to the harmonic cutoffs (black dashed lines)}, while (c-f) focus on
spectra below the ionization energy (vertical red lines). \textcolor{black}{$h_{\kappa}^{\prime}$ \textcolor{black}{(interband)}
mechanism: orange (1.05, 3.15 \textgreek{m}m) and purple (2.1, 4.2 \textgreek{m}m)
diamonds. $h_{\kappa}^{\prime\prime}$ \textcolor{black}{(intraband)}
mechanism: green (1.05, \textcolor{black}{3.15} \textgreek{m}m) and blue (2.1, 4.2 \textgreek{m}m) circles.} \label{fig:Comparison-of-harmonic}
\vspace{-12pt}}
\end{figure}

Although links between the quantum optical channels in Eqs. (\ref{eq:hA}) and (\ref{eq:hB}) and Lewenstein's work are 
established, there is a broader connection between quantum optical and semiclassical models. By using properties of 
Fourier transforms and integration by parts (see supplement Ref. \citep{Supplement}) Eqs. (\ref{eq:hA}) and (\ref{eq:hB}) 
can be rewritten as,
\begin{align}
h_{\kappa}^{\prime}= & \frac{i\eta_{\kappa}\mathbf{e}_{\kappa}}{\hbar}e^{-i\mathbf{k}\mathbf{x}}\int_{-\infty}^{\infty}dte^{i\omega_{k}t}\mathbf{j}_{er}(t),\label{eq:hAJer}\\
h_{\kappa}^{\prime\prime}\approx & \frac{i\eta_{k}\mathbf{e}_{{\kappa}}}{\hbar}e^{-i\mathbf{k}\mathbf{x}}\int_{-\infty}^{\infty}dte^{i\omega_{k}t}\mathbf{j}_{ra}(t),\label{eq:hBJra}
\end{align}
where,
$\mathbf{j}_{er}(t)=\vert e\vert\dot{\mathbf{x}}(t)$ and $\mathbf{j}_{ra}(t)=\int d^{3}p\mathbf{v}(\mathbf{p}_{t})\vert 
b_{\mathbf{p}}(t)\vert^{2}$ represent the time-dependent interband and intraband currents previously studied semiclassically 
as mechanisms of HHG in solids \citep{thorpe2023,vampa2014,vampa2017}. 
Here, we define $\dot{\mathbf{x}}(t)=\partial_{t}\mathbf{x}(t),$ as the time dependent dipole velocity, 
$\mathbf{v}(\mathbf{p}_{t})=\boldsymbol{\nabla}_{\mathbf{p}}(\mathbf{p}^{2}/2m)\vert_{\mathbf{p}=\mathbf{p}_{t}} 
=\mathbf{p}_{t}/m$ as the continuum velocity, and $\vert b_{\mathbf{p}}(t)\vert^{2}$ as the electronic continuum 
population with continuum amplitude $b_{\mathbf{p}}(t)=ie^{-iS(t)}\int_{-\infty}^{t}dt^{\prime}\Omega(t^{\prime}) 
e^{iS(t^{\prime})}$, of the ground-continuum wavefunction as in \citep{lewenstein1994}. 

We observe that the quantum optical results of Eqs. (\ref{eq:hAJer}) and (\ref{eq:hBJra}) consist of semiclassical
currents times a factor with \textcolor{black}{$\eta_{\kappa}\propto\omega_{k}^{-1/2}$} resulting from vacuum fluctuations. \textcolor{black}{ So far, no studies have examined the relation between these two mechanisms in atoms, which is analogous to interband and intraband HHG in solids.} 

To clarify how the interband current $\mathbf{j}_{er}(t)=\vert e\vert\dot{\mathbf{x}}(t)$ and intraband current $\mathbf{j}_{ra}(t)=\int d^{3}p\mathbf{v}(\mathbf{p}_{t})\vert b_{\mathbf{p}}(t)\vert^{2}$ defined for gases relate to the theory of solids \citep{vampa2014,vampa2015}, note that in solids,  the interband current is generated by polarization 
accumulation due to dipole coupling of valence and conduction bands. Technically, the momentum $\mathbf{p}$ is replaced by 
crystal momentum $\mathbf{K}$, the band-velocity is defined as $v(\mathbf{K}) = \boldsymbol{\nabla}_{\mathbf{K}} \varepsilon(\mathbf{K})$, and $\varepsilon$ is the bandgap between valence and conduction band. Next, the dipole is replaced 
by the dipole transition moment between valence and conduction band. \textcolor{black}{In gases the intraband term related to the Brunel mechanism \citep{brunel90}.} In a solid, as the band velocity is in general
a nonlinear function of $\mathbf{K}$, intraband HHG is driven by ionization and band velocity nonlinearities \citep{vampa2014,vampa2015}. 

We explore the two HHG mechanisms in the quantum optical context, as indicated in Eqs. (\ref{eq:hA}) and (\ref{eq:hB}), by 
examining their influence on harmonic photon yield across varying wavelengths. The total photon yield is given by,
\begin{align}
\langle\hat{n}\rangle & =\sum_{s=1,2}\frac{V}{(2\pi)^{3}}\int d^{3}k\vert h_{\kappa}\vert^{2}.
\end{align}
By assuming the quantization volume $V\rightarrow\infty$, that $\mathbf{p}_{t}\parallel k_{z}$ and that one of the 
polarization basis vectors is contained in the plane of $\mathbf{k}$ and $\mathbf{p}_{t}$, we find that the photon 
expectation value variation with harmonic frequency, and the total yield about a given harmonic $N$ are given by,
\begin{align}
\frac{d\langle\hat{n}\rangle}{d\omega_{k}} & =\frac{V\omega_{k}^{2}}{3\pi{}^{2}c^{3}}\vert\bar{h}_{\kappa}\vert^{2},\label{eq:ndiff}
\end{align}
\begin{align}
\langle\hat{n}\rangle_{N} & =\frac{V}{3\pi{}^{2}c^{3}}\int_{\left(N-1/2\right)\omega_{0}}^{\left(N+1/2\right)\omega_{0}}d\omega_{k}\omega_{k}^{2}\vert\bar{h}_{\kappa}\vert^{2},\label{eq:yield}
\end{align}
respectively (see supplement Ref. \citep{Supplement}). Here, $\bar{h}_{\kappa}=h_{\kappa}\csc\theta$ is independent of the polar angle $\theta$ with wavevector $\mathbf{k}$. In addition, note that the factor $V$ cancels out with a factor of 
$1/V$ from $\eta_{\kappa}^{2}$ contained in $\vert\bar{h}_{\kappa}\vert^{2}$.

We now investigate HHG in argon
gas driven by a strong \textcolor{black}{linearly polarized laser} field with a Gaussian temporal envelope
(full width at half maximum of 5 cycles) and a peak intensity of $3.5\times10^{14}\text{ W}/\text{cm}^{2}$.
The total photon yield is calculated as a function of harmonic order
(Eq. (\ref{eq:yield})) for the two mechanisms described in Eqs. (\ref{eq:hAJer})
and (\ref{eq:hBJra}), using driving wavelengths of 1.05 \textgreek{m}m,
2.1 \textgreek{m}m, 3.05 \textgreek{m}m, and 4.2 \textgreek{m}m, and
a simple Coulomb correction factor for ionization \citep{ammosov1986}.

Figures \ref{fig:Comparison-of-harmonic}(a) and (b) show the full harmonic spectra \textcolor{black}{up to the harmonic cutoffs (black dashed lines), determined by the cutoff energy, $E_{\text{max}}=E_0+3.17U_{p}$, where $U_{p}\approx\vert e\vert^2 \mathcal{F}_{0}^2/4 m \omega_0^2$ is the ponderomotive energy given a driving frequency $\omega_0$ and maximum amplitude ${\mathcal{F}}_{0}$  \citep{lewenstein1994,corkum1993}.} Panels (c) through (f) focus
on the low-order harmonics \textcolor{black}{with dashed lines indicating the harmonic orders corresponding to the 
ionization energy for each wavelength}.

Observing these spectra, the interband quantum
optical channel, Eq. (\ref{eq:hA}), is dominant throughout most of the spectra, especially in the HH plateaux,
with cutoffs consistent with semiclassical atomic gas results \citep{wahlstrom1993,chintalwad2024,lewenstein1994,majidi2024,corkum1993,jin2012,schafer1993}. 

In our argon model gas, the lowest order harmonics
are driven by the intraband mechanism of Eq. (\ref{eq:hB}), with
intraband dominance increasing with wavelength. The intraband dominance
noticeably occurs for more harmonics as wavelength is increased.

\begin{figure}[tb]
\textcolor{black}{\includegraphics{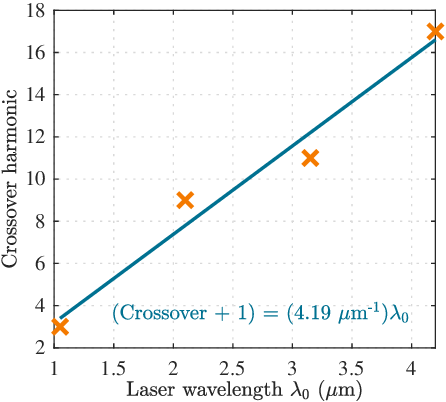}\caption{Comparison of the crossover harmonic (orange crosses), defined as
the highest odd harmonic for which the $h_{\kappa}^{\prime\prime}$ \textcolor{black}{(intraband)} mechanism (Eq. (\ref{eq:hBJra})) is greater or
near to equal to the $h_{\kappa}^{\prime}$ \textcolor{black}{(interband)} mechanism
(Eq. (\ref{eq:hAJer})) for wavelengths as in FIG. \ref{fig:Comparison-of-harmonic}.
A teal linear trendline and its corresponding equation are also plotted.\label{fig:Linear}}
}
\end{figure}
To compare the quantum optical intraband and interband
analogous mechanisms as a function of the driving laser's wavelength,
we plot the \textquotedbl crossover harmonic\textquotedbl{} (the
highest harmonic where the $h_{\kappa}^{\prime\prime}$ spectrum is
greater than or equal to the $h_{\kappa}^{\prime}$ spectrum) in Fig.
\ref{fig:Linear} based on Fig. \ref{fig:Comparison-of-harmonic}.
A linear trend of one plus the crossover harmonic with laser wavelength
is observed, amounting to a constant \textcolor{black}{photon} energy of 5.195 eV. This allows for predicting where the intraband mechanism will dominate, enabling experiments in gas 
to explore and tune the dominant mechanism.

In conclusion, we demonstrated that, in gases, quantizing the electromagnetic field naturally produces mechanisms analogous to
semiclassical interband and intraband HHG in gases and solids. The intraband HHG term in gases is driven by the ionization 
nonlinearity, as the continuum does not have any nonlinear properties, unlike the valence and conduction bands in solids. 
\textcolor{black}{The quantum optical model accounts for vacuum fluctuations which result in an additional factor, the vacuum electric
field, modifying magnitude and wavelength scaling of previous classical-light, quantum-matter theories.}
We found that interband analogous mechanisms dominate the HH plateau in atoms, while intraband analogous mechanisms dominate low order 
harmonics. \textcolor{black}{The crossover between the two regimes relates to a constant photon energy independent of the 
wavelength of the driving laser.} 
% Our work complements semiclassical strong field theories of HHG in that it allows close-to quantitative modeling of HHG. 

%\newpage
%\FloatBarrier

\end{document}

\title{A unified treatment of high-harmonic generation in gases and solids: supplemental material}
\author{\textcolor{black}{A. Thorpe}\textsuperscript{\textcolor{black}{1}}}
\email{athor087@uottawa.ca}

\author{N. Boroumand\textsuperscript{1}}
\author{G. Bart\textsuperscript{1}}
\author{\textcolor{black}{L. Wang}\textsuperscript{\textcolor{black}{1}}}
\author{\textcolor{black}{G. Vampa}\textsuperscript{\textcolor{black}{1,2}}}
\author{\textcolor{black}{T. Brabec}\textsuperscript{\textcolor{black}{1}}}
\affiliation{\textsuperscript{1}Department of Physics, University of Ottawa, Ontario
K1N 6N5, Canada}
\affiliation{\textsuperscript{2}National Research Council of Canada, Ottawa, Ontario
K1A 0R6, Canada}
\date{\today}

\maketitle
In this supplemental document we provide greater detail as to how
we derive the photon wavefunction and corresponding coherent state
parameters for interband-like and intraband-like mechanisms. Further,
we provide some detail on the calculation of expectation values, in
particular the harmonic photon yield. We provide other information
in the appendices: the calculation of $\sigma_{{\!\kappa}}(t)$,
the calculation of the quantum vacuum phase, the filter we use in
our model that suppresses long electron trajectories, and Fourier
transform identities that we utilize.

\section{Derivation of the photon wavefunction\label{sec:derivationwf}}

\noindent We start in the Schrödinger picture, operators time independent,
and wavefunction time dependent. The electric field and vector potential
operators are defined as 
\begin{subequations}
\label{ea} 
\begin{align}
\hat{\mathbf{F}}(\mathbf{x}) & =i\sum_{{\kappa}}\eta_{k}\omega_{k}\mathbf{e}_{{\kappa}}\left(\hat{a}_{{\kappa}}e^{i\mathbf{k}\mathbf{x}}-\hat{a}_{{\kappa}}^{\dagger}e^{-i\mathbf{k}\mathbf{x}}\right)\label{vece}\\
\hat{\mathbf{A}}(\mathbf{x}) & =\sum_{{\kappa}}\eta_{k}\mathbf{e}_{{\kappa}}\left(\hat{a}_{{\kappa}}e^{i\mathbf{k}\mathbf{x}}+\hat{a}_{{\kappa}}^{\dagger}e^{-i\mathbf{k}\mathbf{x}}\right)\text{,}\label{veca}
\end{align}
\end{subequations}
 compare Eqs. (2.119), (2.120) in Ref. \citep{gerry05}. Here, $\mathbf{k}=k\mathbf{n}$
runs over all wavevectors, $\mathbf{n}$ represents the unit vector
along $\mathbf{k}$, $k=\omega_{k}/c$, and $s=1 \text{ or } 2$ is the light
polarization index; to minimize indices, we define the multi-index
${\kappa}\equiv(\mathbf{k},s)$; further, $\mathbf{e}_{{\kappa}}$
specifies the direction of polarization, and hat denotes operator
quantities. Finally, $\eta_{k}^{2}=\hbar/(2\omega_{k}\varepsilon_{0}V)$
and $\varepsilon_{0}$ is the vacuum permittivity.

Our goal is to solve the Schrödinger equation 
\begin{align}
i\hbar\partial_{t}\vert\Psi(t)\rangle=\hat{H}\vert\Psi(t)\rangle=\left(\hat{H}_{m}+\hat{H}_{f}+\hat{H}_{i}\right)\vert\Psi(t)\rangle\mathrm{,}\label{schroed}
\end{align}
where 
\begin{subequations}
\label{Hs} 
\begin{align}
\hat{H}_{m} & =\hat{\mathbf{p}}^{2}/(2m)+V(\mathbf{r})\label{Hm}\\
\hat{H}_{f} & =\sum_{{\kappa}}\hbar\omega_{k}\hat{a}_{{\kappa}}^{\dagger}\hat{a}_{{\kappa}}\label{Hf}\\
\hat{H}_{i} & =\mathbf{r}|e|\hat{\mathbf{F}}(\mathbf{x})\label{Hi}
\end{align}
\end{subequations}
 are the matter, field, and interaction Hamiltonian; $m$ is the electron
mass; we distinguish between local coordinates of a single atom $\mathbf{r}$
and global coordinates $\mathbf{x}$ that vary over the whole material;
$\hat{\mathbf{p}}$, $\mathbf{r}$ are electron momentum and position
relative to the global center of mass $\mathbf{x}$ of the atom.

Eq. (\ref{schroed}) is solved by utilizing the Ansatz 
\begin{align}
 & \vert\Psi(t)\rangle=\vert0\rangle\otimes\vert\phi_{0}(t)\rangle+\int\!\!d^{3}\mkern-2mu p\,\vert\mathbf{p}\rangle\otimes\vert\phi_{\mathbf{p}}(t)\rangle\label{ansatz}
\end{align}
 where $\vert0\rangle$ is the atomic ground state of binding energy
$E_{0}$, and ionized electron continuum state $\vert\mathbf{p}\rangle=1/(2\pi\hbar)^{3/2}\exp((i/\hbar)\mathbf{p}\mathbf{r})$.
Further, photon Fock states $\vert\phi_{0}(t)\rangle$ and $\vert\phi_{\mathbf{p}}(t)\rangle$,
are associated with the ground and continuum states, respectively.

Inserting Ansatz (\ref{ansatz}) into Eq. (\ref{schroed}) and operating
on the resulting equation with functionals $\langle0\vert$ and $\langle\mathbf{p}\vert$
yields two equations of motion for the respective photon wavefunctions,
\begin{subequations}
\label{eqmo} 
\begin{align}
 & i\hbar\partial_{t}\vert\phi_{0}\rangle=(E_{0}+H_{f})\vert\varphi_{0}\rangle+\int\!\!d^{3}\mkern-2mu p\,\mathbf{d}^{*}\!(\mathbf{p})\,\vert e\vert\hat{\mathbf{F}}\,\vert\phi_{\mathbf{p}}\rangle\label{eqmo0}\\
 & i\hbar\partial_{t}\vert\phi_{\mathbf{p}}\rangle\!=\!\Bigl(\frac{\mathbf{p}^{2}}{2m}\!\!+\!H_{f}\!+\!\vert e\vert\hat{\mathbf{F}}(-i\hbar\boldsymbol{\nabla}_{\mathbf{p}})\Bigr)\vert\phi_{\mathbf{p}}\rangle\nonumber \\
 & +\mathbf{d}(\mathbf{p})\,\vert e\vert\hat{\mathbf{F}}\,\vert\phi_{0}\rangle\mathrm{.}\label{eqmop}
\end{align}
\end{subequations}
 In the derivation of Eq. (\ref{eqmo}) we have employed the strong
field approximation; single electron ground and continuum states are approximately orthogonal, $\langle\mathbf{p}\vert0\rangle=\langle0\vert\mathbf{p}\rangle=0$,
and the effect of the Coulomb potential on the continuum electron
is assumed to be negligible compared to the intense laser field. Further,
the single electron transition dipole moment is defined as $\mathbf{d}(\mathbf{p})=\langle\mathbf{p}\vert\mathbf{r}\vert0\rangle=d_{0}\mathbf{p}/(p^{2}+2mE_{0})^{3}$
\citep{brabec22,salpeter77}.

Next, we transform Eqs. (\ref{eqmo}) into velocity the gauge by inserting
$\vert\phi{}_{l}\rangle=\hat{U}_{v}\vert\phi{}_{l}^{\prime}\rangle$,
$l=0,\mathbf{p}$ with unitary operator $\hat{U}_{v}=\exp(i({\vert e\vert}/\hbar)\mathbf{r}\hat{\mathbf{A}})$.
The resulting equations are 
\begin{subequations}
\label{eqv} 
\begin{align}
i\hbar\partial_{t}\vert\phi_{0}^{\prime}\rangle & =(E_{0}+H_{f})\vert\phi_{0}^{\prime}\rangle\nonumber \\
 & +\int\!\!d^{3}\mkern-2mu p\,\mathbf{d}^{*}\!(\mathbf{p}+|e|\hat{\mathbf{A}})\,\vert e\vert\hat{\mathbf{F}}\,\vert\phi_{\mathbf{p}}^{\prime}\rangle\label{eqv0}\\
i\hbar\partial_{t}\vert\phi_{\mathbf{p}}^{\prime}\rangle & \!=\!\Bigl(\frac{1}{2m}(\mathbf{p}\!+\!|e|\hat{\mathbf{A}})^{2}\!+\!H_{f}\Bigr)\vert\phi_{\mathbf{p}}^{\prime}\rangle\nonumber\\
 & +\mathbf{d}(\mathbf{p}+|e|\hat{\mathbf{A}})\,\vert e\vert\hat{\mathbf{F}}\vert\phi_{0}^{\prime}\rangle\mathrm{.}\label{eqvp}
\end{align}
\end{subequations}
 For the derivation of Eq. (\ref{eqmo}) the following relations have
been used; ${\hat{U}_{v}^{\dagger}}f(\mathbf{p})\hat{U}_{v}=f(\mathbf{p+|e|\hat{\mathbf{A}}})$;
${\hat{U}_{v}^{\dagger}}H_{f}\hat{U}_{v}=H_{f}-|e|\mathbf{x}\hat{\mathbf{F}}$;
and ${\color{black}{}{\hat{U}_{v}^{\dagger}}\hat{\mathbf{F}}\hat{U}_{v}=\hat{\mathbf{F}}}$.
For the second relation it is helpful to use $[\hat{a},\exp(\beta\hat{a}^{\dagger})]=\beta\exp(\beta\hat{a}^{\dagger})$
and $[\exp(\beta\hat{a}),\hat{a}^{\dagger}]=\beta\exp(\beta\hat{a})$
\citep{loudon83}.

For the next step we need coherent states. A coherent state can be
expressed in terms of the displacement operator $\vert\alpha(t)\rangle=\hat{D}_{\alpha}(t)\vert{v}\rangle$,
where $\vert{v}\rangle$ is the vacuum state over all
modes, $\vert{v}\rangle=\prod_{{\kappa}}\vert{v}_{{\kappa}}\rangle$,
and the displacement operator is defined as ${\hat{D}_{\alpha}(t)}=\prod_{{\kappa}}{\hat{D}(\alpha_{{\kappa}}(t))}$;
for a single mode, 
\begin{align}
{\hat{D}(\alpha_{{\kappa}}(t))}=\exp\left(\alpha_{{\kappa}}\hat{a}_{{\kappa}}^{\dagger}{\normalcolor {\normalcolor {\color{black}{{\normalcolor e}^{{\normalcolor -i\omega_{k}t}}}}}}-\alpha_{{\kappa}}^{*}\hat{a}_{{\kappa}}{{\normalcolor e^{i\omega_{k}t}}}\right)\mathrm{.}\label{Dks}
\end{align}
Note that we are in the Schrödinger picture, where operators are time
independent. However, in the above we use the displacement operator
to describe a photon wavefunction state which is time dependent. Pulling
the time dependence into the coefficient, we define $\alpha_{{\kappa}}(t)=\alpha_{{\kappa}}e^{-i\omega_{k}t}$.
With Eq. (\ref{Dks}), the coherent state can be expressed as $\vert\alpha_{{\kappa}}(t)\rangle={\hat{D}(\alpha_{{\kappa}}(t))}\vert v_{{\kappa}}\rangle$.

As the driving laser is intense, it makes sense to model it via a
multi-mode coherent state, $\vert\phi_{0}^{\prime}(t)\rangle=\vert\alpha(t)\rangle+\vert\phi_{0}^{\prime\prime}(t)\rangle$.
The weak quantum state $\vert\phi_{0}^{\prime\prime}(t)\rangle$ accounts
for weak perturbations to the pump beam and for the radiation produced
during HHG. We assume a set of quantum modes for
the intense laser ${\kappa}\in\{c\}$ and one for the quantum
state ${\kappa}\in\{q\}$. The modes of the quantum state
are assumed to not overlap with the intense laser modes, $\{c\}\cap\{q\}=\emptyset$.
For now we assume that $\vert\phi_{0}^{\prime\prime}(t_{0})\rangle=0$.
As a result, we can say ${\hat{D}_{\alpha}^{\dagger}(t)}\vert\phi_{0}^{\prime}(t)\rangle=\vert\phi_{0}^{\prime\prime}(t)\rangle$,
as ${\hat{D}_{\alpha}^{\dagger}(t)}\vert\alpha(t)\rangle=\vert{v}\rangle$.
The displacement operator ${\hat{D}_{\alpha}(t)}=\prod_{{\kappa}\in\{c\}}{\hat{D}(\alpha_{{\kappa}}(t))}$
runs over the modes of the intense laser.

In case there is a quantum field $\vert\varepsilon(t)\rangle$ that
overlaps with the intense laser modes we would need to carry along
${\hat{D}_{\alpha}^{\dagger}}(t)\vert\varepsilon(t)\rangle$
in the following transformation. 

In order to remove the coherent state from the
wavefunction, we make the replacement $\vert\phi_{l}^{\prime}(t)\rangle={\hat{D}_{\alpha}(t)}\vert\phi_{l}^{\prime\prime}(t)\rangle$
($l=0,\mathbf{p}$) in Eq. (\ref{eqv}). The vector potential transforms
as ${\hat{D}_{\alpha}^{\dagger}}(t)\hat{\mathbf{A}}{\hat{D}_{\alpha}}(t)=\tilde{\mathbf{A}}(t)=\boldsymbol{\mathcal{A}}(t)+\hat{\mathbf{A}}$,
and the electric field becomes ${\hat{D}_{\alpha}^{\dagger}}(t)\hat{\mathbf{F}}{\hat{D}_{\alpha}(t)}=\tilde{\mathbf{F}}(t)=\boldsymbol{\mathcal{F}}(t)+\hat{\mathbf{F}}$.
The quantum fields are defined as in Eq. (\ref{ea}) and the classical
fields are given by 
\begin{subequations}
\label{fields} 
\begin{align}
\boldsymbol{\mathcal{F}}_{\!}(t) & \!=\!i\!\!\!\sum_{{\kappa}\in\{c\}}\!\!\!\eta_{k}\omega_{k}\mathbf{e}_{{\kappa}}\!\left(\alpha_{{\kappa}}e^{i(\mathbf{k}\mathbf{x}-\omega_{k}t)}\!-\!\alpha_{{\kappa}}^{*}e^{-i(\mathbf{k}\mathbf{x}-\omega_{k}t)}\right)\label{ecl}\\
\boldsymbol{\mathcal{A}}_{\!}(t) & =\!\!\!\sum_{{\kappa}\in\{c\}}\!\!\eta_{k}\mathbf{e}_{{\kappa}}\left(\alpha_{{\kappa}}e^{i(\mathbf{k}\mathbf{x}-\omega_{k}t)}+\alpha_{{\kappa}}^{*}e^{-i(\mathbf{k}\mathbf{x}-\omega_{k}t)}\right)\text{.}\label{acl}
\end{align}
\end{subequations}
 %
\begin{comment}
For the derivation of the classical fields, the displacement operator
is split using the Baker-Campbell-Haussdorff formula \citep{loudon83}
and then commuted with the field operators via the relations given
below Eq. (\ref{eqv}).
\end{comment}
{} Further, we have to work out $\partial_{t}{\hat{D}_{\alpha}(t)}$
and ${\hat{D}_{\alpha}^{\dagger}(t)}\hat{H}_{f}{\hat{D}_{\alpha}(t)}$
where the second expression gives an additional term that cancels
the time derivative of the displacement operator
\citep{kaminer21}.

The resulting equations of motion are 
\begin{subequations}
\label{eqvd} 
\begin{align}
 & i\hbar\partial_{t}\vert\phi_{0}^{\prime\prime}\rangle\!=\!(E_{0}+H_{f})\vert\phi_{0}^{\prime\prime}\rangle\!+\!\!\!\int d^{3}\mkern-2mu p\,\mathbf{d}^{*}\!\bigl(\mathbf{p}\!+\!|e|\tilde{\mathbf{A}}(t)\bigr)\times\nonumber \\
 & \times\vert e\vert\tilde{\mathbf{F}}(t)\,\vert\phi_{\mathbf{p}}^{\prime\prime}\rangle\label{eqvd0}\\
 & i\hbar\partial_{t}\vert\phi_{\mathbf{p}}^{\prime\prime}\rangle\!=\!\Bigl(\frac{1}{2m}(\mathbf{p}\!+\!|e|\tilde{\mathbf{A}}(t))^{2}\!+\!H_{f}\Bigr)\vert\phi_{\mathbf{p}}^{\prime\prime}\rangle\nonumber \\
 & +\mathbf{d}\bigl(\mathbf{p}+|e|\tilde{\mathbf{A}}(t)\bigr)\,\vert e\vert\tilde{\mathbf{F}}(t)\vert\phi_{0}^{\prime\prime}\rangle\mathrm{.}\label{eqvdp}
\end{align}
\end{subequations}

To put classical and quantum fields on equal footing, we transform
to the interaction picture; we chose $\vert\phi_{l}^{\prime\prime}\rangle=\hat{U}_{i}\vert\varphi_{l}\rangle$,
$l=0,\mathbf{p}$ with $\hat{U}_{i}=\exp(-i\hat{H}_{f}t{/\hbar})$.
The resulting equations are 
\begin{subequations}
\label{eqvdi} 
\begin{align}
 & i\hbar\partial_{t}\vert\varphi_{0}\rangle\!=\!E_{0}\vert\varphi_{0}\rangle\!+\!\int\!\!d^{3}\mkern-2mu p\,\mathbf{d}^{*}\!\bigl(\mathbf{p}\!+\!|e|{\tilde{\mathbf{A}}(t)}\bigr)\,\vert e\vert{\tilde{\mathbf{F}}(t)}\,\vert\varphi_{\mathbf{p}}\rangle\label{eqvdi0}\\
 & i\hbar\partial_{t}\vert\varphi_{\mathbf{p}}\rangle=\Bigl(\frac{1}{2m}(\mathbf{p}+|e|\tilde{\mathbf{A}}(t))^{2}\Bigr)\vert\varphi_{\mathbf{p}}\rangle\nonumber \\
 & +\mathbf{d}\bigl(\mathbf{p}+|e|\tilde{\mathbf{A}}(t)\bigr)\,\vert e\vert\tilde{\mathbf{F}}(t)\vert\,\varphi_{0}\rangle\mathrm{.}\label{eqvdip}
\end{align}
\end{subequations}
 In the interaction picture, the operators $\hat{a}_{{\kappa}}\exp(-i\omega_{k}t)$
and $\hat{a}_{{\kappa}}^{\dagger}\exp(i\omega_{k}t)$ become
time dependent, and as a consequence the field and vector potential
operators, $\tilde{\mathbf{F}}(t)=\boldsymbol{\mathcal{F}}(t)+\hat{\mathbf{F}}(t)$,
$\tilde{\mathbf{A}}(t)=\boldsymbol{\mathcal{A}}(t)+\hat{\mathbf{A}}(t)$.
For the derivation of Eq. (\ref{eqvdi}) the following relations are
helpful; $\exp(i\omega t\hat{a}^{\dagger}\hat{a})\hat{a}\exp(-i\omega t\hat{a}^{\dagger}\hat{a})=\hat{a}\exp(-i\omega t)$
and $\exp(i\omega t\hat{a}^{\dagger}\hat{a})\hat{a}^{\dagger}\exp(-i\omega t\hat{a}^{\dagger}\hat{a})=\hat{a}^{\dagger}\exp(i\omega t)$
\citep{loudon83}.

The quantum fields clearly present a perturbation to the intense pump
field. As a result, we introduce the following approximations: (i)
we expand the dipole moments with regard to the operator arguments
to first order, $\mathbf{d}(\mathbf{p}_{t}+\vert e\vert\hat{\mathbf{A}}(t))\approx\mathbf{d}(\mathbf{p}_{t})+\vert e\vert[(\hat{\mathbf{A}}(t)\boldsymbol{\nabla}_{\mathbf{p}})\mathbf{d}(\mathbf{p})]_{\mathbf{p}_{t}}$,
where $\mathbf{p}_{\mkern-1.5mu t}=\mathbf{p}+|e|\boldsymbol{\mathcal{A}}(t)$.
While we do not carry along the first order contribution in this work,
we keep in mind that it is there. (ii) we approximate $(\mathbf{p}+|e|\tilde{\mathbf{A}}(t))^{2}\approx{\mathbf{p}_{t}^{2}}+2{\mathbf{p}_{t}}|e|\hat{\mathbf{A}}(t)$.
With these approximations, Eqs. (\ref{eqvdi}) become 
\begin{subequations}
\label{eqvdi1} 
\begin{align}
i\hbar\partial_{t}\vert\varphi_{0}\rangle & =E_{0}\vert\varphi_{0}\rangle+\int\!\!d^{3}\mkern-2mu p\,\mathbf{d}^{*}\!(\mathbf{p}_{t})\,\vert e\vert\tilde{\mathbf{F}}(t)\,\vert\varphi_{\mathbf{p}}\rangle\label{eqvdi01}\\
i\hbar\partial_{t}\vert\varphi_{\mathbf{p}}\rangle & =\left(\frac{1}{2m}\mathbf{p}_{t}^{2}+\frac{|e|}{m}\mathbf{p}_{t}\hat{\mathbf{A}}(t)\right)\vert\varphi_{\mathbf{p}}\rangle\nonumber \\
 & +\mathbf{d}(\mathbf{p}_{t})\,\vert e\vert\tilde{\mathbf{F}}(t)\,\vert\varphi_{0}\rangle\mathrm{.}\label{eqvdip1}
\end{align}
\end{subequations}

Eqs. (\ref{eqvdi1}) can be integrated analytically. We start with
Eq. (\ref{eqvdip1}) which contains an operator term 
\begin{align}
-i\frac{|e|}{\hbar m}\mathbf{p}_{\mkern-1.5mu t}\hat{\mathbf{A}}(t) & =\sum_{{\kappa}\in\{q\}}\overset{\bm{.}}{\sigma}_{{\!\kappa}}(t)\hat{a}_{{\kappa}}^{\dagger}-\overset{\bm{.}}{\sigma}_{{\!\kappa}}^{*}(t)\hat{a}_{{\kappa}}\nonumber \\
\overset{\bm{.}}{\sigma}_{{\!\kappa}}(t) & =-i\frac{|e|\eta_{k}}{\hbar m}\bigl(\mathbf{e}_{{\kappa}}\mathbf{p}_{\mkern-1.5mu t}\bigr)e^{-i(\mathbf{k}\mathbf{x}-\omega_{k}t)},\label{sigma}
\end{align}
where the dot on $\overset{\bm{.}}{\sigma}_{{\!\kappa}}(t)$
indicates the time derivative $\overset{\bm{.}}{\sigma}_{{\!\kappa}}(t)=\partial_{t}\sigma_{{\kappa}}(t)$.
The exponent of this operator is clearly a displacement operator,
\begin{align}
\exp\left(-i\frac{|e|}{\hbar m}\mathbf{p}_{\mkern-1.5mu t}\hat{\mathbf{A}}(t)\right)=\prod_{{\kappa}\in\{q\}}\!\!{\hat{D}(\overset{\bm{.}}{\sigma}_{{\!\kappa}}(t))},\label{Dsig}
\end{align}
where the displacement operator was defined above
Eq. (\ref{Dks}).

For the integration of Eq. (\ref{eqvdip1}) the following relation
is useful,
\begin{align}
 \partial_{t}\exp(\hat{B}(t))=(\partial_{t}\hat{B}+(1/2)[\hat{B}(t),\partial_{t}\hat{B}])\exp(\hat{B}(t)),   
\end{align}
as long as the commutator gives a c-number. Equation (\ref{eqvdip1})
is integrated by using the Ansatz 
\begin{align}
\vert\varphi_{\mathbf{p}}\rangle & ={\hat{D}_{\sigma}(t^{\prime})}\,\exp\!\left(i\!\!\int_{{t_{0}}}^{t'}\!\!\!g'dt''\right)\vert\tilde{\varphi}_{\mathbf{p}}\rangle\textrm{,}\label{opans}
\end{align}
where, $\hat{D}_{\sigma}(t)=\prod_{{\kappa}\in\{q\}}{\hat{D}(\sigma_{{\!\kappa}}(t))}$,

\begin{align}
\sigma_{{\!\kappa}}(t)= & \sigma_{{\!\kappa}}(t_{0},t)=\int_{{t_{0}}}^{t}\overset{\bm{.}}{\sigma}_{{\!\kappa}}dt'\nonumber \\
= & {\frac{\vert e\vert E_{v}}{\hbar}}\overline{\sigma}_{\!{\kappa}}(t)e^{-i(\mathbf{k}\mathbf{x}-\omega_{k}t)}\\
{\overline{\sigma}_{\!{\kappa}}(t)=} & {\frac{-i}{m\omega_{k}}}{\int_{t_{0}}^{t}\mathbf{e}_{{\kappa}}\mathbf{p}_{\mkern-1.5mu t^{\prime}}e^{-i\omega_{k}(t-t^{\prime})}dt',}\label{eq:sigmabara}
\end{align}
 $t_{0}$ is the initial time, and $E_{v}=\eta_{k}\omega_{k}$
is the vacuum electric field amplitude. An evaluation
of this integral is given in Appendix \ref{sec:sigmacalc}.

The purpose of the exponential c-number term in the Ansatz Eq. (\ref{opans})
is to cancel the commutator that arises from the time derivative of
the operator part of the Ansatz above.

We insert Ansatz (\ref{opans}) into Eq. (\ref{eqvdip1}). Evaluating
the time derivative, as prescribed above, a commutator results that
needs to be evaluated using relation (\ref{sigma}), 
\begin{align}
 & i{g'(t)}=\frac{1}{2}\left[-i\frac{|e|}{\hbar m}\int_{t_{0}}^{t}dt'\mathbf{p}_{\mkern-1.5mu t'}\hat{\mathbf{A}}(t'),-i\frac{|e|}{\hbar m}\mathbf{p}_{\mkern-1.5mu t}\hat{\mathbf{A}}(t)\right]=\nonumber \\
 & =i\sum_{{\kappa}}\mathrm{Im}\left[\overset{\bm{.}}{\sigma}_{{\!\kappa}}^{*}(t)\sigma_{{\!\kappa}}(t)\right]\label{opcommu}
\end{align}

After elimination of the operator term, the resulting equation can
be directly integrated to give 
\begin{align}
 & \vert\tilde{\varphi}_{\mathbf{p}}\rangle=-\frac{i}{\hbar}\int_{t_{0}}^{t}dt'\exp\!\left(i\int_{t_{0}}^{t'}\!\!\Bigl(\frac{\mathbf{p}_{\mkern-1.5mu \tau}^{2}}{2m\hbar}-{g^{\prime}(\tau)}\Bigr)\,d\tau\right)\nonumber \\
 & \times\mathbf{d}(\mathbf{p}_{\mkern-1.5mu t^{\prime}})\,\vert e\vert\tilde{\mathbf{F}}(t'){\hat{D}_{\sigma}(t^{\prime})}\vert\tilde{\varphi}_{0}(t')\rangle\label{tildphi}
\end{align}

Transformation back to $\vert\varphi_{\mathbf{p}}\rangle$ we encounter
a product of displacement operators that need to be combined to a
single displacement operator, 
\begin{align}
 & \prod_{{\kappa}}{\hat{D}(\sigma_{{\!\kappa}}(t))}{\hat{D}^{\dagger}(\sigma_{{\!\kappa}}(t^{\prime}))}=\nonumber \\
= & {\hat{D}_{\sigma}(t^{\prime},t)}\exp\bigl(ig^{\prime\prime}(t',t)\bigr)\mathrm{,}\label{combD}
\end{align}
with ${\hat{D}_{\sigma}(t^{\prime},t)}=\prod_{{\kappa}}{\hat{D}(\sigma_{{\!\kappa}}(t^{\prime},t))}$.
Further,
\begin{align}
 & ig^{\prime\prime}(t',t)=-\frac{1}{2}\left(\frac{|e|}{\hbar m}\right)^{2}\left[\int_{t_{0}}^{t}\!\!\!d\tau\mathbf{p}_{\mkern-1.5mu \tau}\hat{\mathbf{A}}_{j}(\tau),\int_{t_{0}}^{t'}\!\!\!d\tau\mathbf{p}_{\mkern-1.5mu \tau}\hat{\mathbf{A}}(\tau)\right]\nonumber \\
 & =i\sum_{{\kappa}}\mathrm{Im}\left[\sigma_{{\!\kappa}}(t')\,\sigma_{{\!\kappa}}^{*}(t)\right]\textrm{.}\label{combD2}
\end{align}
Here we have used the relation $\hat{D}(\alpha)\hat{D}(\beta)=\hat{D}(\alpha+\beta)\exp(i\mathrm{Im}(\alpha\beta^{*}))$
\citep{gerry05}. We define 
\begin{align}
g(t',t)=\int_{t'}^{t}{g'(\tau)}d\tau+g^{\prime\prime}(t',t)\label{g}
\end{align}

Putting everything together we find 
\begin{align}
 & \vert\varphi_{\mathbf{p}}\rangle=-\frac{i|e|}{\hbar}\int_{{t_{0}}}^{t}dt'\exp\left(-\frac{i}{2m\hbar}\int_{t'}^{t}\mathbf{p}_{\mkern-1.5mu \tau}^{2}d\tau+ig(t',t)\right)\nonumber \\
 & \times\mathbf{d}(\mathbf{p}_{t'})\,\mathbf{\tilde{F}}(t')\,{\hat{D}_{\sigma}(t^{\prime},t)}\,\vert\varphi_{0}\rangle.\label{tildvarphip}
\end{align}

Finally, we integrate Eq. (\ref{eqvdi0}) and insert Eq. (\ref{tildvarphip})
to obtain 
\begin{align}
 & \vert\varphi_{0}(t)\rangle\!=\!{{\normalcolor \vert\varphi_{0}(t_{0})\rangle}}\!-\!\frac{|e|^{2}}{\hbar^{2}}\int_{t_{0}}^{t}\!\!\!\!dt'\!\!\int\!\!d^{3}\mkern-2mu p\,\mathbf{d}^{*}\!(\mathbf{p}_{\mkern-1.5mu t^{\prime}}\!)\,{{\color{black}\tilde{\mathbf{F}}(t^{\prime})}}\!\!\int_{{t_{0}}}^{t'}\!\!\!dt''\nonumber \\
 & {{\normalcolor \tilde{\mathbf{F}}(t^{\prime\prime})}}\,\mathbf{d}(\mathbf{p}_{\mkern-1.5mu t^{\prime\prime}}\!)\,e^{-iS(t^{\prime\prime},t^{\prime})+ig(t^{\prime\prime},t^{\prime})}{\hat{D}_{\sigma}(t^{\prime\prime},t^{\prime})}\vert\varphi_{0}(t'')\rangle\textrm{}\label{tildvarphi0}
\end{align}
with 
\begin{align}
S({t})={\frac{1}{\hbar}}{\int_{t_{0}}^{t}}\!\left(\frac{1}{2m}\mathbf{p}_{\mkern-1.5mu \tau}^{2}+E_{0}\right)d\tau\mathrm{.}\label{S}
\end{align}
the classical action, and $S(t^{\prime},t)=S(t)-S(t^{\prime})$.

Eq. (\ref{tildvarphi0}) can be simplified by defining the right-hand
side as,
\begin{align}
 & -\int\!\!d^{3}\mkern-2mu p\!\int_{t_{0}}^{t}\!\!dt^{\prime}\hat{c}_{\mathbf{p}}^{\dagger}(t^{\prime})\!\!\int_{t_{0}}^{t^{\prime}}\!\!\!dt^{\prime\prime}\hat{c}_{\mathbf{p}}(t^{\prime\prime})\bigl\vert\varphi_{0}(t^{\prime\prime})\bigr\rangle=\nonumber \\
 & -\frac{|e|^{2}}{\hbar^{2}}\int\!\!d^{3}\mkern-2mu p\,\,\mathbf{d}^{*}\!(\mathbf{p}_{t^{\prime}})\int_{t0}^{t^{\prime}}\!\!\!dt^{\prime\prime}\mathbf{d}(\mathbf{p}_{t}\!)\nonumber \\
 & \times\left(\!\boldsymbol{\mathcal{F}}(t^{\prime})\boldsymbol{\mathcal{F}}(t^{\prime\prime})\!+\!\hat{\mathbf{F}}(t^{\prime})\boldsymbol{\mathcal{F}}(t^{\prime\prime})\!+\!\boldsymbol{\mathcal{F}}(t^{\prime})\hat{\mathbf{F}}(t^{\prime\prime})\!\right)\nonumber \\
 & \times\!{\hat{D}_{\sigma}(t^{\prime\prime},t^{\prime})}\exp\left(-iS(t^{\prime\prime},t^{\prime})\right)\bigl\vert\varphi_{0}\bigr\rangle(t^{\prime\prime}),\label{hatpi}
\end{align}
\begin{align}
 & \hat{c}_{\mathbf{p}}(t^{\prime\prime})=\tilde{\Omega}(t^{\prime\prime}){\hat{D}_{\sigma}^{\dagger}}(t^{\prime\prime})\exp\left(iS(t^{\prime\prime})\right)\mathrm{.}\label{hatpinew}
\end{align}
Here, $\tilde{\Omega}(t^{\prime\prime})=(\vert e\vert/\hbar)\mathbf{d}(\mathbf{p}_{t^{\prime\prime}})\tilde{\mathbf{F}}(t^{\prime\prime})\bigr)=\Omega(t)+\hat{\Omega}(t)$
is a generalized Rabi frequency with the quantum Rabi frequency $\hat{\Omega}(t)=\sum_{\kappa}\hat{\Omega}_{\kappa}(t)$.
Further, the term $g(t'',t')$ is omitted as can be analyzed in Appendix
\ref{sec:Quantum-vacuum-phase}.

Using Eq. (\ref{hatpinew}) we can rewrite Eq.
(\ref{tildvarphi0}) to arrive at the following integro-differential
equation,
\begin{align}
\partial_{t}\vert\varphi_{0}(t)\rangle=-\int\!\!d^{3}\mkern-2mu p\,\hat{c}_{\mathbf{p}}^{\dagger}(t)\!\!\int_{t_{0}}^{t}\!\!\!dt^{\prime}\,\hat{c}_{\mathbf{p}}(t^{\prime})\,\vert\varphi_{0}(t^{\prime})\rangle.\label{dtldphi0}
\end{align}
In the weak depletion limit, $\vert\varphi_{0}\rangle$ can be pulled
out of the integral on the right hand side of Eq. (\ref{dtldphi0})
by integration by parts; higher order terms are neglected. The resulting
differential equation of motion is 
\begin{align}
\partial_{t}\vert\varphi_{0}(t)\rangle\approx\left(\partial_{t}\hat{\frak{h}}(t)\right)\vert\varphi_{0}(t)\rangle\mathrm{.}\label{eqmo-1}
\end{align}
Integration of Eq. (\ref{eqmo-1}) by the method of Magnus and Fer
\citep{wilcox67} results in 
\begin{align}
 & \vert\varphi_{0}(t)\rangle\approx\exp\left(\hat{\frak{h}}(t)\right)\vert\varphi_{0}(t_{0})\rangle\label{eq:phi0appr}\\
 & \hat{\frak{h}}=-\int\!\!d^{3}\mkern-2mu p\int_{t_{0}}^{t}\!\!\!dt^{\prime}\,\hat{c}_{\mathbf{p}}^{\dagger}(t^{\prime})\int_{t_{0}}^{t^{\prime}}\!\!\!dt^{\prime\prime}\,\hat{c}_{\mathbf{p}}(t^{\prime\prime})\mathrm{.}\label{tldphi0}
\end{align}

The next order term in the Magus Fer expansion, 
\begin{align}
\exp\left(\int_{t_{0}}^{t}\!\!dt^{\prime}\!\!\int_{t_{0}}^{t}\!\!dt^{\prime\prime}\left[\partial_{t^{\prime}}\hat{\frak{h}},\partial_{t^{\prime\prime}}\hat{\frak{h}}\right]\right)\mathrm{,}\label{magnusho}
\end{align}
is small and is neglected, see the discussion of the zero- and one-photon
operator terms in the next section.

\section{Zero- and one-photon operator terms}

\noindent The operator $\exp(\hat{\frak{h}}(t))$ is not unitary due
to the coupling between ground state and the continuum photon wavefunction.
In order to isolate the unitary part of the wavefunction, $\hat{\frak{h}}(t)$
is split into an anti-Hermitian part plus small remainder. For that,
we keep only zero-, and one-operator terms $\hat{\frak{h}}\approx\frak{h}^{(0)}+\hat{\frak{h}}^{(1)}$,
respectively and neglect the rest. This is reasonable, as $\vert\sigma_{\kappa}\vert,\vert\Omega_{\kappa}\vert\ll1$.

The expansion of $\hat{\frak{h}}$ is done by first expanding $\hat{c}_{\mathbf{p}}$
in Eq. (\ref{hatpinew}) up to one-photon operator terms, $\hat{c}_{\mathbf{p}}\approx\hat{c}_{\mathbf{p}}^{(0)}+\hat{c}_{\mathbf{p}}^{(1)}$
with 
\begin{align}
 & \hat{c}_{\mathbf{p}}^{(0)}(t)=\Omega(t)e^{iS(t)}\nonumber \\
 & \hat{c}_{\mathbf{p}}^{(1)}(t)=\sum_{\kappa}\left(\hat{\Omega}_{\kappa}(t)-\Omega(t)\hat{\sigma}_{\kappa}(t)\right)e^{iS(t)},\label{eq:hatc}
\end{align}
where, $\hat{\sigma}_{\kappa}(t)=\sigma_{\kappa}(t)\hat{a}_{\kappa}^{\dagger}-\sigma_{\kappa}^{*}(t)\hat{a}_{\kappa}$.

From this $\hat{\frak{h}}=\hat{\frak{h}}^{(0)}+\hat{\frak{h}}^{(1)}$
in relation (\ref{tldphi0}) is determined up to first order with
\begin{subequations}
\label{hath} 
\begin{align}
 & \hat{\frak{h}}^{(0)}=-\int\!\!d^{3}\mkern-2mu p\int_{t_{0}}^{t}\!\!\!dt^{\prime}\left(\hat{c}_{\mathbf{p}}^{(0)*}\hat{C}_{\mathbf{p}}^{(0)}\right)\!(t^{\prime})\label{hath0}\\
 & \hat{\frak{h}}^{(1)}=-\int\!\!d^{3}\mkern-2mu p\int_{t_{0}}^{t}\!\!\!dt^{\prime}\left(\hat{c}_{\mathbf{p}}^{(0)*}\hat{C}_{\mathbf{p}}^{(1)}+\hat{c}_{\mathbf{p}}^{(1)\dagger}\hat{C}_{\mathbf{p}}^{(0)}\right)\!(t^{\prime})\label{hath1}
\end{align}
\end{subequations}
 Here, we have introduced $\hat{C}_{\mathbf{p}}^{(j)}(t)=\int_{-\infty}^{t}dt^{\prime}\hat{c}_{\mathbf{p}}^{(j)}(t^{\prime})$
($j=0,1$) for the inner time integral. Further, HHG terms should
have the photon operator term outside of the inner time integral,
i.e. they should all be represented by small letter terms $\hat{c}_{\mathbf{p}}^{(1)}$.
To achieve this, we perform integration by parts of the terms in Eqs.
(\ref{hath1}) containing $\hat{C}_{\mathbf{p}}^{(1)}$. For example,
$\int^{t}\!dt^{\prime}\hat{c}_{\mathbf{p}}^{(0)*}\hat{C}_{\mathbf{p}}^{(1)}=-\int^{t}\!dt^{\prime}\hat{c}_{\mathbf{p}}^{(1)}\hat{C}_{\mathbf{p}}^{(0)*}+R$,
where $R$ is a small residual term.

Applying the above steps to Eq. ({\ref{hath}}) yields 
\begin{subequations}
\label{hathm} 
\begin{align}
 & \hat{\frak{h}}^{(0)}=-\int\!\!d^{3}\mkern-2mu p\int_{t_{0}}^{t}\!\!\!dt^{\prime}\left(\hat{c}_{\mathbf{p}}^{(0)*}\hat{C}_{\mathbf{p}}^{(0)}\right)\!(t^{\prime})\label{hathm0}\\
 & \hat{\frak{h}}^{(1)}\approx\int\!\!d^{3}\mkern-2mu p\int_{t_{0}}^{\infty}\!\!\!dt^{\prime}\left(\hat{c}_{\mathbf{p}}^{(1)}\hat{C}_{\mathbf{p}}^{(0)*}-\hat{c}_{\mathbf{p}}^{(1)\dagger}\hat{C}_{\mathbf{p}}^{(0)}\right)\!(t^{\prime})\label{hathm1}
\end{align}
\end{subequations}
 where we also have let $t\rightarrow\infty$ in the HH term (\ref{hathm1}).

Next, we evaluate the individual terms in Eq. (\ref{hathm}) by inserting
Eq. (\ref{eq:hatc}). For Eq. (\ref{hathm0}) one obtains 
\begin{align}
 & \frak{h}^{(0)}(t)=-\int_{t_{0}}^{t}\!\!\!dt^{\prime}\gamma(t^{\prime})=-\int_{t_{0}}^{t}\!\!\!dt^{\prime}\!\int\!d^{3}\mkern-2mu p\,\Gamma_{\mathbf{p}}(t^{\prime})\nonumber \\
 & \Gamma_{\mathbf{p}}(t)=\Omega^{*}(t)e^{-iS(t)}\!\!\int_{t_{0}}^{t}\!\!\!dt^{\prime}\,\Omega(t^{\prime})\,e^{iS(t^{\prime})}\mathrm{.}\label{0op}
\end{align}
Here, $\gamma(t)$ is a complex rate, and $\gamma+\gamma^{*}$ gives
the optical field ionization rate from ground state to continuum.

The one photon operator contribution (\ref{hathm1}) is worked out
to be $\hat{\frak{h}}^{(1)}=\sum_{\kappa}\hat{\frak{h}}_{\kappa}^{(1)}$
with 
\begin{align}
\hat{\frak{h}}_{\kappa}^{(1)}=h_{\kappa}\hat{a}_{\kappa}^{\dagger}-h_{\kappa}^{*}\hat{a}_{\kappa}\mathrm{.}\label{oneop1}
\end{align}
The next order term in the Magnus Fer expansion (\citep{wilcox67})
is a small phase term that does not contribute to HHG and can be neglected.
Therefore, assuming the initial state is vacuum, Eq. (\ref{tldphi0})
for HHG gives a coherent state with the HHG coefficient given by 
\begin{align}
h_{\kappa} & =e^{-i\mathbf{k}\mathbf{x}}\!\!\int_{-\infty}^{\infty}\!\!\!\!dt^{\prime}e^{i\omega_{k}t^{\prime}}H_{k}(t^{\prime})=\tilde{H}_{k}e^{-i\mathbf{k}\mathbf{x}}\label{hhg}\\
H_{k} & =\Biggl\{\frac{\vert e\vert E_{v}}{\hbar}\bigl(\mathbf{e}_{\kappa}\mathbf{x}(t^{\prime})\bigr)-\int\!\!d^{3}\mkern-2mu p\,\overline{\sigma}_{\kappa}(t^{\prime})\left(\Gamma_{\mathbf{p}}(t^{\prime})+\mathrm{c.c}\right)\Biggr\}\mathrm{}\nonumber 
\end{align}
with 
\begin{align}
\mathbf{x}(t)= & 2\text{Re}\left[i\int d^{3}p\;\mathbf{d}^{*}(\mathbf{p}_{t})e^{-iS(t)}\int_{-\infty}^{t}dt^{\prime}\Omega(t^{\prime})e^{iS(t^{\prime})}\right].\label{eq:lewdip}
\end{align}
The first term in Eq. (\ref{hhg}) relates to previous work \citep{lewenstein94},
where $\mathbf{x}(t)$ is the expectation value of the dipole moment
driving HHG. Note that the subscript in $H_{k}$ is chosen $k$ to
indicate that it does not depend on the HHG wavevector, but only on
the frequency. It is also assumed that harmonic and laser polarization
vectors are parallel, so that the polarization index can be neglected
as well. 

Finally, it should be noted that Eq. (\ref{hhg}) does not contain
any explicit terms that account for dephasing between the ground state
and the continuum. Without any kind of dephasing mechanism implemented
in the model, the spectrum becomes too noisy to easily distinguish
differing odd harmonics. To counteract this, we include in our numerical
simulations a filter as in Appendix \ref{sec:Long-electron-trajectory}.
This is based on the fact that HHG is dominated
by the shortest trajectories, where the time between ionization and
harmonic emission is within a half-cycle. 

\section{Connection to interband and intraband currents in solids \label{sec:Connection-to-interband}}

From Eq. (\ref{hhg}) we see that the coherent quantum optical HHG
parameter, $h_{\kappa}$, consists of a sum of two parts,\textcolor{black}{
\begin{subequations}
\textcolor{black}{\label{hAB}
\begin{align}
h_{\kappa}^{\prime}= & \frac{\vert e\vert E_{v}}{\hbar}\mathbf{e}_{\kappa}\int_{-\infty}^{\infty}dte^{i(\omega_{k}t-\mathbf{k}\mathbf{x})}\mathbf{x}(t),\label{eq:hA}\\
h_{\kappa}^{\prime\prime}= & -{\frac{\vert e\vert E_{v}}{\hbar}}\int_{-\infty}^{\infty}dte^{i(\omega_{k}t-\mathbf{k}\mathbf{x})}\int d^{3}p\bar{\sigma}_{\kappa}(t)\left[\Gamma_{\mathbf{p}}(t)+\text{c.c.}\right].\label{eq:hB}
\end{align}
}
\end{subequations}
We will demonstrate that they can be connected to the theory of interband
and intraband currents in solids, respectively \citep{vampa2014}.
This thus bridges HHG theory of gases and solids as well as quantum
optical and semiclassical theories.}

To derive these connections, we assume a linearly polarized classical driving field that is of the
form, $\boldsymbol{\mathcal{F}}(t)=\mathcal{\boldsymbol{\mathcal{F}}}_{0}f(t)\sin(\omega_{0}t)$, where $\mathcal{\boldsymbol{\mathcal{F}}}_{0}$ is the field vector of maximum amplitude, $\omega_0$ is its frequency, and
$f(t)$ is a slowly varying temporal envelope. We aim to rewrite Eqs. (\ref{hAB}) to
eliminate integer powers of the harmonic frequency factors $\omega_{k}$,
isolating the role of vacuum field fluctuations $\eta_{\kappa}\propto\omega_{k}^{-1/2}$.
This involves using the differentiation (Eq. (\ref{eq:hB})) or integration
(Eq. (\ref{eq:hA})) properties of Fourier transforms (see Appendix \ref{sec:Integration-and-differentiation}).

In order to facilitate the demonstration of these connections, let
us write here some correspondences between atomic gas systems and
two-band solid \citep{vampa2014} systems:
\begin{subequations}
\begin{align}
\mathbf{p}\rightarrow & \mathbf{k}\in\text{BZ}\label{eq:ptok}\\
E_{0}+\frac{\mathbf{p}^{2}}{2m}\rightarrow & \varepsilon(\mathbf{k})=E_{c}(\mathbf{k})-E_{v}(\mathbf{k})\\
\mathbf{v}(\mathbf{p})= & \boldsymbol{\nabla}_{\mathbf{p}}\left(E_{0}+\frac{\mathbf{p}^{2}}{2m}\right)=\frac{\mathbf{p}}{m}\nonumber \\
\rightarrow & \mathbf{v}(\mathbf{k})=\hbar^{-1}\boldsymbol{\nabla}_{\mathbf{k}}\varepsilon(\mathbf{k})\\
\mathbf{d}(\mathbf{p})\rightarrow & \mathbf{d}(\mathbf{k})=i\int d^{3}\mathbf{x}u_{c,\mathbf{k}}^{*}(\mathbf{x})\boldsymbol{\nabla}_{\mathbf{k}}u_{v,\mathbf{k}}(\mathbf{x}).\label{eq:dk}
\end{align}
\end{subequations}
That is, ionized electronic momentum $\mathbf{p}$ is replaced with
the crystal wavevector $\mathbf{k}$ (or crystal momentum $\hbar\mathbf{k}$)
within the Brillouin zone (BZ), where all physics for $\mathbf{k}\not\in\text{BZ}$
can be remapped to. Further, the atomic ground state and electronic
continuum energies are replaced with the valence band $E_{v}(\mathbf{k})$
and the conduction band $E_{c}(\mathbf{k})$, respectively. $\varepsilon(\mathbf{k})$
describes the energy difference between these two bands and is often
more convenient to use for calculations. The continuum electronic
velocity $\mathbf{v}(\mathbf{p})=\mathbf{p}/m$ is analogous to the
band difference velocity, or more simply, the band velocity $\mathbf{v}(\mathbf{k})$.
Note that both the valence and conduction bands have velocities associated
with them due to their curvatures in $\mathbf{k}$-space. This is
not the case with an atomic system where only the ionized electron
motion contributes to the system's velocity. Finally, the ground-continuum
transition dipole $\mathbf{d}(\mathbf{p})$ in a gas is replaced with
the valence-conduction transition dipole $\mathbf{d}(\mathbf{k})$,
as in Eq. (\ref{eq:dk}), where, $u_{m,\mathbf{k}}(\mathbf{x}),m=v,c$
are Bloch functions that correspond to periodic parts of crystal wavefunctions.

\subsection{Interband current}

Let us start by rewriting Eq. (\ref{eq:hA}) in terms of $\eta_{\kappa}$
as,
\begin{align}
h_{\kappa}^{\prime}= & \frac{\vert e\vert\eta_{\kappa}\omega_{k}}{\hbar}\mathbf{e}_{\kappa}\int_{-\infty}^{\infty}dte^{i(\omega_{k}t-\mathbf{k}\mathbf{x})}\mathbf{x}(t).\label{eq:hAeta}
\end{align}
To eliminate the $\omega_{k}$ factor we first rewrite $\mathbf{x}(t)=\int_{-\infty}^{t}dt^{\prime}\dot{\mathbf{x}}(t^{\prime})$
so that,
\begin{align}
h_{\kappa}^{\prime}= & \frac{\vert e\vert\eta_{\kappa}\omega_{k}}{\hbar}\mathbf{e}_{\kappa}\int_{-\infty}^{\infty}dte^{i(\omega_{k}t-\mathbf{k}\mathbf{x})}\int_{-\infty}^{t}dt^{\prime}\dot{\mathbf{x}}(t^{\prime}).\label{eq:hAeta-1}
\end{align}
We now apply the integral Fourier transform identity (see Eq. (\ref{eq:intprop})
in Appendix \ref{sec:Integration-and-differentiation}), ignoring
the DC term, to obtain,
\begin{align}
h_{\kappa}^{\prime}= & \frac{i\vert e\vert\eta_{\kappa}}{\hbar}\mathbf{e}_{\kappa}\int_{-\infty}^{\infty}dte^{i(\omega_{k}t-\mathbf{k}\mathbf{x})}\dot{\mathbf{x}}(t).\label{eq:hAeta-1-1}
\end{align}
Outside of the Fourier exponent $e^{i\omega_{k}t}$, this form now
only scales with $\omega_{k}$ through vacuum fluctuations via $\eta_{\kappa}\propto\omega_{k}^{-1/2}$.

Using Eqs. (\ref{eq:ptok})-(\ref{eq:dk}), when
we replace the dipole $\mathbf{d}(\mathbf{p})$ and action $S(t)$
in $\mathbf{x}(t)$ of Eq. (\ref{eq:lewdip}) by the solid state
analogues, then we get a connection betwenn $h_{\kappa}^{\prime}$
and the quasiclassical interband current in solids \citep{vampa2014},
\begin{align}
h_{\kappa}^{\prime}= & \frac{i\eta_{\kappa}\mathbf{e}_{\kappa}}{\hbar}e^{-i\mathbf{k}\mathbf{x}}\mathbf{J}_{er}(\omega_{k}),\label{eq:hpjer}
\end{align}
where, 
\begin{align}
\mathbf{J}_{er}(\omega_{k})= & \int_{-\infty}^{\infty}dte^{i\omega_{k}t}\mathbf{j}_{er}(t),
\end{align}
represents the interband current in the frequency domain and
$\mathbf{j}_{er}(t)=\vert e\vert\dot{\mathbf{x}}(t)$ is in the
time domain. Thus, vacuum fluctuations $\eta_{\kappa}\propto\omega_{k}^{-1/2}$
connect the coherent state parameter $h_{\kappa}^{\prime}$ to the
interband current that can be derived semiclassically. 

\subsection{Intraband current}

Consider Eq. (\ref{eq:hB}), it can rewritten in an alternative way
as,

\begin{align}
h_{\kappa}^{\prime\prime} & =-{\frac{\vert e\vert E_{v}}{\hbar}}\int d^{3}p\int_{-\infty}^{\infty}dte^{-i(\mathbf{k}\mathbf{x}-\omega_{k}t)}\bar{\sigma}_{\kappa}(t)\partial_{t}\vert b_{\mathbf{p}}(t)\vert^{2},
\end{align}
where \textcolor{black}{$\vert b_{\mathbf{p}}(t)\vert^{2}$ is the
electronic continuum population with continuum amplitude $b_{\mathbf{p}}(t)=ie^{-iS(t)}\int_{-\infty}^{t}dt^{\prime}\Omega(t^{\prime})e^{iS(t^{\prime})}$,
of the atom-continuum wavefunction as in }\citep{lewenstein94}\textcolor{black}{.
}Of note is that by examining Eq. (\ref{tildvarphip})
we see that, up to a phase term, $b_{\mathbf{p}}(t)$ represents the classical field part of the ground-continuum photon wavefunction overlap $\langle\varphi_{0}\vert\varphi_{\mathbf{p}}\rangle$.
\textcolor{black}{We made use of the fact that $\partial_{t}\vert b_{\mathbf{p}}(t)\vert^{2}=\Gamma_{\mathbf{p}}(t)+\mathrm{c.c}$
according to Eq. (\ref{0op}).}

Inserting $\bar{\sigma}_{\kappa}(t)$, as evaluated
in Eq. (\ref{eq:sigkap-1}) of Appendix \ref{sec:sigmacalc}, and
assuming for the harmonics in question that $\omega_{k}^{2}\gg\omega_{0}^{2}$,
we obtain,
\begin{align}
h_{\kappa}^{\prime\prime} & \approx-\frac{|e|\eta_{k}\mathbf{e}_{{\kappa}}}{\hbar m}\int d^{3}p\int_{-\infty}^{\infty}dte^{-i(\mathbf{k}\mathbf{x}-\omega_{k}t)}\nonumber \\
\times & \left\lbrace \frac{\mathbf{p}_{t}}{\omega_{k}}+\frac{|e|}{\omega_{k}^{2}}\left[\frac{\omega_{0}^{2}}{\omega_{k}}\boldsymbol{\mathcal{A}}(t)-i\boldsymbol{\mathcal{F}}(t)\right]\right\rbrace \partial_{t}\vert b_{\mathbf{p}}(t)\vert^{2}.
\end{align}

Next, we can use the relation $\partial_{t}\left(\boldsymbol{\mathcal{A}}(t)\vert b_{\mathbf{p}}(t)\vert^{2}\right)=-\boldsymbol{\mathcal{F}}(t)\vert b_{\mathbf{p}}(t)\vert^{2}+\boldsymbol{\mathcal{A}}(t)\partial_{t}\vert b_{\mathbf{p}}(t)\vert^{2}$
in the above to obtain,
\begin{align}
h_{\kappa}^{\prime\prime} & =-\frac{|e|\eta_{k}\mathbf{e}_{{\kappa}}}{\hbar m}\int d^{3}p\int_{-\infty}^{\infty}dte^{-i(\mathbf{k}\mathbf{x}-\omega_{k}t)}\Bigg\lbrace\frac{\mathbf{p}_{t}}{\omega_{k}}\partial_{t}\vert b_{\mathbf{p}}(t)\vert^{2}\nonumber \\
+ & \vert e\vert\frac{\omega_{0}^{2}}{\omega_{k}^{3}}\boldsymbol{\mathcal{A}}(t)\partial_{t}\vert b_{\mathbf{p}}(t)\vert^{2}-\vert e\vert\frac{i}{\omega_{k}^{2}}\boldsymbol{\mathcal{F}}(t)\partial_{t}\vert b_{\mathbf{p}}(t)\vert^{2}\Bigg\rbrace.
\end{align}
This can be modified further by using the relation $\boldsymbol{\mathcal{A}}(t)\partial_{t}\vert b_{\mathbf{p}}(t)\vert^{2}=\partial_{t}\left(\boldsymbol{\mathcal{A}}(t)\vert b_{\mathbf{p}}(t)\vert^{2}\right)+\boldsymbol{\mathcal{F}}(t)\vert b_{\mathbf{p}}(t)\vert^{2}$
and applying the differentiation rule of Fourier transforms, as in
Eq. (\ref{eq:diffprop}) of Appendix \ref{sec:Integration-and-differentiation},
on the term with $\partial_{t}\left(\boldsymbol{\mathcal{A}}(t)\vert b_{\mathbf{p}}(t)\vert^{2}\right)$
to obtain,
\begin{align}
h_{\kappa}^{\prime\prime} & =\frac{i|e|\eta_{k}\mathbf{e}_{{\kappa}}}{\hbar m}\int d^{3}p\int_{-\infty}^{\infty}dte^{-i(\mathbf{k}\mathbf{x}-\omega_{k}t)}\nonumber \\
\times & \Bigg\lbrace\frac{i\mathbf{p}_{t}}{\omega_{k}}\partial_{t}\vert b_{\mathbf{p}}(t)\vert^{2}+\vert e\vert\frac{\omega_{0}^{2}}{\omega_{k}^{2}}\left(\boldsymbol{\mathcal{A}}(t)\vert b_{\mathbf{p}}(t)\vert^{2}\right)\nonumber \\
+ & i\vert e\vert\frac{\omega_{0}^{2}}{\omega_{k}^{3}}\boldsymbol{\mathcal{F}}(t)\vert b_{\mathbf{p}}(t)\vert^{2}+\frac{\vert e\vert}{\omega_{k}^{2}}\boldsymbol{\mathcal{F}}(t)\partial_{t}\vert b_{\mathbf{p}}(t)\vert^{2}\Bigg\rbrace.
\end{align}

Next we assume that $i\vert e\vert(\omega_{0}^{2}/\omega_{k}^{3})\boldsymbol{\mathcal{F}}(t)\vert b_{\mathbf{p}}(t)\vert^{2}$
is much smaller than the other terms in braces so that we can use,
\begin{gather}
h_{\kappa}^{\prime\prime}\approx\frac{i|e|\eta_{k}\mathbf{e}_{{\kappa}}}{\hbar m}\int d^{3}p\int_{-\infty}^{\infty}dte^{-i(\mathbf{k}\mathbf{x}-\omega_{k}t)}\Bigg\lbrace\frac{i\mathbf{p}_{t}}{\omega_{k}}\partial_{t}\vert b_{\mathbf{p}}(t)\vert^{2}\nonumber \\
+\vert e\vert\frac{\omega_{0}^{2}}{\omega_{k}^{2}}\left(\boldsymbol{\mathcal{A}}(t)\vert b_{\mathbf{p}}(t)\vert^{2}\right)+\frac{\vert e\vert}{\omega_{k}^{2}}\boldsymbol{\mathcal{F}}(t)\partial_{t}\vert b_{\mathbf{p}}(t)\vert^{2}\Bigg\rbrace.
\end{gather}
With this result we can follow a similar procedure as before by utilizing
the relation, valid in the slowly varying envelope approximation,
$\partial_{t}\left(\boldsymbol{\mathcal{F}}(t)\vert b_{\mathbf{p}}(t)\vert^{2}\right)\approx\omega_{0}^{2}\boldsymbol{\mathcal{A}}(t)\vert b_{\mathbf{p}}(t)\vert^{2}+\boldsymbol{\mathcal{F}}(t)\partial_{t}\vert b_{\mathbf{p}}(t)\vert^{2}$.
Using this relation and applying the differentiation rule of Fourier
transforms (Eq. (\ref{eq:diffprop}) of Appendix \ref{sec:Integration-and-differentiation})
on the resulting term with $\partial_{t}\left(\boldsymbol{\mathcal{F}}(t)\vert b_{\mathbf{p}}(t)\vert^{2}\right)$,
we obtain,
\begin{align}
h_{\kappa}^{\prime\prime} & =\frac{i|e|\eta_{k}\mathbf{e}_{{\kappa}}}{\hbar m}\int d^{3}p\int_{-\infty}^{\infty}dte^{-i(\mathbf{k}\mathbf{x}-\omega_{k}t)}\nonumber \\
\times & \Bigg\lbrace\frac{i\mathbf{p}_{t}}{\omega_{k}}\partial_{t}\vert b_{\mathbf{p}}(t)\vert^{2}-i\frac{\vert e\vert}{\omega_{k}}\boldsymbol{\mathcal{F}}(t)\vert b_{\mathbf{p}}(t)\vert^{2}\Bigg\rbrace.
\end{align}

Finally, by using the relation $\partial_{t}\left(\mathbf{p}_{t}\vert b_{\mathbf{p}}(t)\vert^{2}\right)=-\vert e\vert\boldsymbol{\mathcal{F}}(t)\vert b_{\mathbf{p}}(t)\vert^{2}+\mathbf{p}_{t}\partial_{t}\vert b_{\mathbf{p}}(t)\vert^{2}$,
we arrive at,

\begin{align}
h_{\kappa}^{\prime\prime} & =\frac{i|e|\eta_{k}\mathbf{e}_{{\kappa}}}{\hbar m}\int d^{3}p\int_{-\infty}^{\infty}dte^{-i(\mathbf{k}\mathbf{x}-\omega_{k}t)}\nonumber \\
\times & \frac{i}{\omega_{k}}\partial_{t}\left(\mathbf{p}_{t}\vert b_{\mathbf{p}}(t)\vert^{2}\right).
\end{align}
Once again applying the differentiation rule of Fourier transforms
(Eq. (\ref{eq:diffprop}) of Appendix \ref{sec:Integration-and-differentiation})
we arrive at,
\begin{align}
h_{\kappa}^{\prime\prime} & =\frac{i|e|\eta_{k}\mathbf{e}_{{\kappa}}}{\hbar}\int d^{3}p\int_{-\infty}^{\infty}dte^{-i(\mathbf{k}\mathbf{x}-\omega_{k}t)}\mathbf{v}(\mathbf{p}_{t})\vert b_{\mathbf{p}}(t)\vert^{2},\label{eq:hkapsemifinal}
\end{align}
where we made use of the electronic continuum velocity relation $\mathbf{v}(\mathbf{p})=\mathbf{p}/m$. 

Eq. (\ref{eq:hkapsemifinal}) can now be rewritten in the compact
form,
\begin{align}
h_{\kappa}^{\prime\prime} & =\frac{i\eta_{k}\mathbf{e}_{{\kappa}}}{\hbar}e^{-i\mathbf{k}\mathbf{x}}\mathbf{J}_{ra}(\omega_{k}),\label{eq:hppjra}
\end{align}
where, 
\begin{align}
\mathbf{J}_{ra}(\omega_{k}) & =\int_{-\infty}^{\infty}dte^{i\omega_{k}t}\mathbf{j}_{ra}(t),\label{eq:jraom}
\end{align}
and,
\begin{align}
\mathbf{j}_{ra}(t) & =\int d^{3}p\mathbf{v}(\mathbf{p}_{t})\vert b_{\mathbf{p}}(t)\vert^{2}.\label{eq:jrat}
\end{align}

In reference to HHG in solids \citep{vampa2014}, we call Eq. (\ref{eq:jrat})
the ``intraband current'' and Eq. (\ref{eq:jraom}) is its harmonic
frequency domain analogue. To see this connection, consider the form
of the intraband current for a two-band crystal model, as in Ref.
\citep{vampa2014}, of
\begin{align}
\mathbf{j}_{ra}(t) & =\sum_{m=c,v}\int_{\overline{\text{BZ}}}\mathbf{v}_{m}\left(\mathbf{K}_{t}\right)n_{m}(\mathbf{K},t)d^{3}\mathbf{K},\label{eq:jrasolid}
\end{align}
where, in SI units, $\mathbf{v}_{m}\left(\mathbf{k}\right)=\hbar^{-1}\boldsymbol{\nabla}_{\mathbf{k}}E_{m}(\mathbf{k})$
and $n_{m}(\mathbf{K},t)$ correspond to conduction ($m=c$) and valence
($m=v$) band velocities and populations, respectively. As can be
seen by Eq. (\ref{eq:jrasolid}), the intraband current is driven
by displacement of charge within the bands, through a combination
of ionization ($n_{m}(\mathbf{K},t)$) and charge motion ($\mathbf{v}_{m}\left(\mathbf{K}_{t}\right)$).
The nonlinearities of these processes lead to HHG that carry a signature
in Eq. (\ref{eq:jraom}). The connection to Eq. (\ref{eq:jrat}) in
gases is apparent as the only difference in form is that the current
only applies to continuum electrons rather than two (or more) bands.
Although, the role of the continuum velocity cannot be neglected in
the gaseous model, without further external conditions that lead to
anharmonicities, it plays a lesser role compared to the nonlinear
ionization term ($\vert b_{\mathbf{p}}(t)\vert^{2}$), and also, band
velocities of solids where anharmonic and often complicated band curvatures
can lead to nonlinear acceleration of charges. 

With the connection to the solid state intraband current established
for Eq. (\ref{eq:jrat}), we go back to Eq. (\ref{eq:hppjra}). We
find that vacuum fluctuations $\eta_{\kappa}\propto\omega_{k}^{-1/2}$
connect the coherent state parameter $h_{\kappa}^{\prime\prime}$
to the intraband current that can be derived semiclassically. 

\section{Expectation values \label{expval}}

\noindent We look at the general expression for an expectation value
of an operator of the form $\hat{\mathcal{O}}=\hat{\mathcal{O}}(\ldots,\hat{a}_{{\kappa}},\hat{a}_{{\kappa}}^{\dagger},\ldots)$.
The wavefunction in the original frame is given
by Eq. (\ref{ansatz}). The derived photon wavefunctions,
$\vert\varphi_{l}\rangle$ with $l=0,\mathbf{p}$, as in Sec. \ref{sec:derivationwf},
relate to $\vert\phi_{l}\rangle$ in Eq. (\ref{ansatz})
through the relation ${\vert\varphi_{l}\rangle={\hat{D}_{\alpha}^{\dagger}}\hat{U}_{v}^{\dagger}\hat{U}_{i}^{\dagger}\vert\phi_{0}\rangle}$.
To determine the expectation value we need to connect it with the
wavefunction in the original frame, 
\begin{align}
 & \vert\Psi\rangle=\hat{U}_{v}{\hat{D}_{\alpha}}\hat{U}_{i}\vert\Psi\rangle_{v}\nonumber \\
 & \vert\Psi\rangle_{v}=\vert0\rangle_{v}\otimes\vert\varphi_{0}(t)\rangle+\int\!\!d^{3}\mkern-2mu p\,\vert\mathbf{p}\rangle_{v}\otimes\vert\varphi_{\mathbf{p}}(t)\rangle\label{trnsbac}
\end{align}
where $\vert0\rangle_{v}={\hat{U}_{v}^{\dagger}}\vert0\rangle$,
$\vert\mathbf{p}\rangle_{v}={\hat{U}_{v}^{\dagger}}\vert\mathbf{p}\rangle$.

From here we can go two ways; transform the wavefunction back to the
Schrödinger and velocity gauge frame, or transforming the operator
$\hat{\mathcal{O}}$ and keeping the wavefunction in the interaction
frame, or any mixture of both. The easiest way seems to transform
$\hat{\mathcal{O}}$ and to keep $\vert\Psi\rangle_{v}$, 
\begin{align}
 & \langle\Psi\vert\hat{\mathcal{O}}(\hat{a}_{{\kappa}},\hat{a}_{{\kappa}}^{\dagger})\vert\Psi\rangle=\text{}_{v}\langle\Psi\vert{\hat{U}_{i}^{\dagger}{\hat{D}_{\alpha}^{\dagger}}\hat{U}_{v}^{\dagger}}\hat{\mathcal{O}}\hat{U}_{v}{\hat{D}_{\alpha}}\hat{U}_{i}\vert\Psi\rangle_{v}\nonumber \\
 & =\text{}_{v}\langle\Psi\vert{\hat{U}_{v}^{\dagger}}\hat{\mathcal{O}}\bigl(\hat{a}_{{\kappa}}e^{-i\omega_{k}t}\!+\!\alpha(t),\hat{a}_{{\kappa}}^{\dagger}e^{i\omega_{k}t}\!+\!\alpha^{*}(t)\bigr)\hat{U}_{v}\vert\Psi\rangle_{v}\nonumber \\
 & \approx\text{}_{v}\langle\Psi\vert{\hat{U}_{v}^{\dagger}}\hat{\mathcal{O}}\bigl(\ldots\hat{a}_{{\kappa}}e^{-i\omega_{k}t},\hat{a}_{{\kappa}}^{\dagger}e^{i\omega_{k}t},\ldots\bigr)\hat{U}_{v}\vert\Psi\rangle_{v}\nonumber \\
 & \approx\text{}_{v}\langle\Psi\vert\hat{\mathcal{O}}\bigl(\ldots\hat{a}_{{\kappa}}e^{-i\omega_{k}t},\hat{a}_{{\kappa}}^{\dagger}e^{i\omega_{k}t},\ldots\bigr)\vert\Psi\rangle_{v}\mathrm{,}\label{comment}
\end{align}
where the classical field part can be neglected if we are only interested
in HHG quantum emission and where ${\hat{U}_{v}^{\dagger}}\hat{\mathcal{O}}\hat{U}_{v}=\hat{\mathcal{O}}$,
as long as $\hat{\mathcal{O}}$ does not depend on the electronic
part of the wavefunction.

Splitting the wavefunction in Eq. (\ref{trnsbac}) into ground state
and continuum part, $\vert\Psi\rangle_{v}=\vert\Psi_{0}\rangle_{v}+\vert\Psi_{c}\rangle_{v}$,
the operator expectation value can be written as 
\begin{align}
\text{}_{v}\langle\Psi\vert\hat{\mathcal{O}}\vert\Psi\rangle_{v}\!= & \text{ }\!\text{}_{v}\langle\Psi_{0}\vert\hat{\mathcal{O}}\vert\Psi_{0}\rangle_{v}\!+\!\text{}_{v}\langle\Psi_{c}\vert\hat{\mathcal{O}}\vert\Psi_{c}\rangle_{v}\nonumber \\
+ & \!\bigl(\text{}_{v}\langle\Psi_{0}\vert\hat{\mathcal{O}}\vert\Psi_{c}\rangle_{v}\!+\!\mathrm{c.c.}\bigr)\mathrm{.}\label{eq:Fexpv}
\end{align}
We focus on $\text{}_{v}\langle\Psi_{0}\vert\hat{\mathcal{O}}\vert\Psi_{0}\rangle_{v}$
which contains the dominant HHG contribution; the resulting probability
wavefunction is very close to the Lewenstein expression for HHG \citep{lewenstein94}.
Therefore, continuum-continuum terms are omitted. Mixed continuum
ground state matrix elements are zero due to the assumed approximate
orthogonality of plane waves and ground state. We have 
\begin{align}
\text{}_{v}\langle\Psi\vert\hat{\mathcal{O}}\vert\Psi\rangle_{v}\approx\text{}_{v}\langle\Psi_{0}\vert\hat{\mathcal{O}}\vert\Psi_{0}\rangle_{v}=\langle\varphi_{0}\vert\hat{\mathcal{O}}\vert\varphi_{0}\rangle\mathbf{.}\label{Fexpv1}
\end{align}

\subsection{High harmonic photon number expectation value}

The harmonic spectrum is determined by the expectation
value of the number operator $\hat{n}=\sum_{\kappa}\hat{a}^\dagger_{\kappa}\hat{a}_\kappa$. 

Let us consider Eq. (\ref{eq:phi0appr}) with only $\hat{\frak{h}}_{\kappa}^{(1)}=h_{\kappa}\hat{a}_{\kappa}^{\dagger}-h_{\kappa}^{*}\hat{a}_{\kappa}$
and the assumption that the initial state is vacuum such that we are
working with a coherent state with single mode harmonic coherent state
parameters $h_{\kappa}$ as defined in Eq. (\ref{hhg}). Following
the convention of Eq. (\ref{Fexpv1}), and noting that the expectation
value for a single mode coherent state $\vert\alpha\rangle$ is $\langle\alpha\vert\hat{n}\vert\alpha\rangle=\vert\alpha\vert^{2}$
\citep{gerry05}, we find,
\begin{align}
\langle\varphi_{0}\vert\hat{n}\vert\varphi_{0}\rangle & =\langle\hat{n}\rangle\nonumber \\
= & \sum_{\kappa}\vert h_{\kappa}\vert^{2}.\label{eq:cohexpsum}
\end{align}
Assuming that the quantization volume $V$ is
much greater than the scale of the laser wavelengths in the problem,
we can switch the summation to integration in Eq. (\ref{eq:cohexpsum}).
As a result, following Ref. \citep{gerry05} we can write,
\begin{align}
\langle\hat{n}\rangle & =\sum_{s=1,2}\frac{V}{(2\pi)^{3}}\int d^{3}k\vert h_{\kappa}\vert^{2},
\end{align}
where the sum over $s=1,2$ corresponds to the light polarization
basis. Note that the volume factor $V$ is cancelled out by a $\eta_{\kappa}^{2}\propto1/V$
contained in $\vert h_{\kappa}\vert^{2}$.

Now note that the magnitude of the wavevector $\mathbf{k}$ can be
linked to the harmonic frequency through the relation $k=\omega_{k}/c$.
Because of this, we can implement harmonic frequencies into the integral
by switching to spherical coordinates. That is,
\begin{align}
\frac{d\langle\hat{n}\rangle}{d\omega_{k}} & =\sum_{s=1,2}\frac{V\omega_{k}^{2}}{(2\pi c)^{3}}\int d\Omega_{\mathbf{k}}\vert h_{\kappa}\vert^{2},
\end{align}
where the integral over $d\Omega_{\mathbf{k}}=\sin\theta d\theta d\phi$
corresponds to the solid angle associated with $\mathbf{k}$.

We next make the following assumptions: 
\begin{enumerate}
\item $\mathbf{p}_{t}\parallel k_{z}$ 
\item Assume a polarization state in the plane of $\mathbf{k}$ and $\mathbf{p}_{t}$.
Then the second polarization state drops out as it is perpendicular
to $\mathbf{p}_{t}$.
\end{enumerate}
With these assumptions made, note that $h_{\kappa}$ contains a dot
product of the form $\mathbf{e}_{{\kappa}}\mathbf{p}_{\mkern-1.5mu t}=-p_{t}\sin\theta$,
and this also applies to $h_{\kappa}^{\prime}$ and $h_{\kappa}^{\prime\prime}$. 

To separate out angular variables, let $h_{\kappa}=\bar{h}_{\kappa}\sin\theta$,
so that,
\begin{align}
\frac{d\langle\hat{n}\rangle}{d\omega_{k}} & =\frac{V\omega_{k}^{2}}{(2\pi c)^{3}}\int d\Omega_{\mathbf{k}}\sin^{2}\theta\vert\bar{h}_{\kappa}\vert^{2}\nonumber \\
 & =\frac{V\omega_{k}^{2}}{(2\pi c)^{3}}\int_{0}^{2\pi}d\phi\int_{0}^{\pi}d\theta\sin^{3}\theta\vert\bar{h}_{\kappa}\vert^{2}\nonumber \\
 & =\frac{V\omega_{k}^{2}}{3\pi{}^{2}c^{3}}\vert\bar{h}_{\kappa}\vert^{2}.
\end{align}
Utilizing this result we can determine the harmonic photon yield centred
about a given harmonic order $N$ as,
\begin{align}
\langle\hat{n}\rangle_{N} & =\frac{V}{3\pi{}^{2}c^{3}}\int_{\left(N-1/2\right)\omega_{0}}^{\left(N+1/2\right)\omega_{0}}d\omega_{k}\omega_{k}^{2}\vert\bar{h}_{\kappa}\vert^{2}.
\end{align}

\appendix

\section{Calculation of $\sigma_{{\!\kappa}}(t)$ in the limit
$t_{0}\rightarrow-\infty$\label{sec:sigmacalc}}

Here we calculate $\sigma_{{\kappa}}(t)={\vert e\vert E_{v}}\overline{\sigma}_{\!{\kappa}}(t)e^{-i(\mathbf{k}\mathbf{x}-\omega_{k}t)}{/\hbar}$
assuming $t_{0}\rightarrow-\infty$, starting from Eq. (\ref{eq:sigmabara}).
To this end we start from the following definition

\begin{equation}
\sigma_{{\kappa}}(t)=-i\frac{|e|\eta_{k}}{\hbar m}\int_{-\infty}^{t}dt^{\prime}\mathbf{p}_{t^{\prime}}\mathbf{e}_{{\kappa}}e^{i(\omega_{k}t^{\prime}-\mathbf{k}\mathbf{x})}.
\end{equation}

We use integration by parts by letting $u=\mathbf{p}_{t^{\prime}}\mathbf{e}_{{\kappa}}$,
$du=-|e|\boldsymbol{\mathcal{F}}(t^{\prime})\mathbf{e_{{\kappa}}}dt^{\prime}$,
$dv=e^{i\omega_{k}t'}dt^{\prime}$ and $v=-ie^{i\omega_{k}t^{'}}/{\omega_{k}}$.
Then we arrive at

\begin{align}
{\sigma_{{\kappa}}(t)}\;  & {\!\!=\!\!-\frac{|e|\eta_{k}\mathbf{e}_{{\kappa}}e^{-i\mathbf{k}\mathbf{x}}}{\hbar\omega_{k}m}\!\left\lbrace \mathbf{p}_{t}e^{i\omega_{k}t}\!\!+\!\!|e|\int_{-\infty}^{t}\!\!\!\!dt^{\prime}\boldsymbol{\mathcal{F}}(t^{\prime})e^{i\omega_{k}t'}\!\right\rbrace .}
\end{align}

Let $\mathbf{I}(t)=\int_{-\infty}^{t}dt^{\prime}\boldsymbol{\mathcal{F}}(t^{\prime})e^{i\omega_{k}t'}$.
Assume that the classical field is linearly polarized and of the
form, $\boldsymbol{\mathcal{F}}(t)=\mathcal{\boldsymbol{\mathcal{F}}}_{0}f(t)\sin(\omega_{0}t)$,
with a slowly varying envelope $f(t)$, driving field maximum amplitude vector $\mathcal{\boldsymbol{\mathcal{F}}}_{0}$ and driving frequency $\omega_0$ so that $\boldsymbol{\mathcal{A}}(t)\approx\boldsymbol{\mathcal{F}}_{0}f(t)\cos\omega_{0}t/\omega_{0}$.
Therefore, it follows that,

\begin{align}
\mathbf{I}(t) & \approx\frac{\boldsymbol{\mathcal{F}}_{0}f(t)}{2i}\int_{-\infty}^{t}dt^{\prime}\left(e^{i(\omega_{k}+\omega_{0})t'}-e^{i(\omega_{k}-\omega_{0})t'}\right).
\end{align}

\begin{align*}
 & =\frac{-\boldsymbol{\mathcal{F}}_{0}f(t)e^{i\omega_{k}t}}{2(\omega_{k}^{2}-\omega_{0}^{2})}\Big(\omega_{k}(e^{i\omega_{0}t}-e^{-i\omega_{0}t})\\
 & -\omega_{0}(e^{i\omega_{0}t}+e^{-i\omega_{0}t})\Big)
\end{align*}

\begin{align*}
 & =\frac{-\boldsymbol{\mathcal{F}}_{0}f(t)e^{i\omega_{k}t}}{(\omega_{k}^{2}-\omega_{0}^{2})}\left(i\omega_{k}\sin\omega_{0}t-\omega_{0}\cos\omega_{0}t\right)
\end{align*}

\begin{align}
 & =\frac{-e^{i\omega_{k}t}}{(\omega_{k}^{2}-\omega_{0}^{2})}\left(i\omega_{k}\boldsymbol{\mathcal{F}}(t)-\omega_{0}^{2}\boldsymbol{\mathcal{A}}(t)\right).
\end{align}

Then we arrive at

\begin{align}
\sigma_{{\kappa}}(t) & =-\frac{|e|\eta_{k}\mathbf{e}_{{\kappa}}}{\hbar m}e^{i\omega_{k}t-i\mathbf{k}\mathbf{x}}\Bigg\lbrace\frac{\mathbf{p}_{t}}{\omega_{k}}\nonumber \\
+ & \frac{|e|}{(\omega_{k}^{2}-\omega_{0}^{2})}\left(\frac{\omega_{0}^{2}}{\omega_{k}}\boldsymbol{\mathcal{A}}(t)-i\boldsymbol{\mathcal{F}}(t)\right)\Bigg\rbrace.
\end{align}
Noting the definition of $\overline{\sigma}_{\!{\kappa}}(t)$
in Eq. (\ref{eq:sigmabara}), it is given by,\textcolor{black}{
\begin{align}
{\bar{\sigma}_{\kappa}(t)\approx} & {-\frac{\mathbf{e}_{{\kappa}}}{m\omega_{k}}\Bigg\lbrace\frac{\mathbf{p}_{t}}{\omega_{k}}+\frac{|e|}{(\omega_{k}^{2}-\omega_{0}^{2})}}\nonumber \\
{\times} & {\left[\frac{\omega_{0}^{2}}{\omega_{k}}\boldsymbol{\mathcal{A}}(t)-i\boldsymbol{\mathcal{F}}(t)\right]\Bigg\rbrace.}\label{eq:sigkap-1}
\end{align}
}

\section{Quantum vacuum phase $g(t^{\prime\prime},t^{\prime})$\label{sec:Quantum-vacuum-phase}}

\label{renorm}

\noindent We evaluate $g(t^{\prime\prime},t^{\prime})$ as defined
in Eqs. (\ref{opcommu}) (\ref{combD2}) and (\ref{g}). We
start with,
\begin{align}
 & {\int_{t^{\prime}}^{t^{\prime\prime}}\!}\!\!g^{\prime}d\tau\!=\!\sum_{\mathbf{k}}\left(\frac{\vert e\vert\eta_{k}}{\hbar m}\right)^{2}\!\!\!\frac{1}{\omega_{k}}\biggl[\int_{t^{\prime\prime}}^{t^{\prime}}\!\!\!\bigl(\mathbf{e}_{{\kappa}}\mathbf{p}_{\mkern-1.5mu \tau}\bigr)^{\!2}\!d\tau\!-\!\frac{1}{\omega_{k}}\bigl(\mathbf{e}_{{\kappa}}\mathbf{p}_{\mkern-1.5mu t_{0}}\bigr)\biggr.\nonumber \\
 & {\biggl.\times\bigl\{\bigl(\mathbf{e}_{{\kappa}}\mathbf{p}_{\mkern-1.5mu t^{\prime\prime}}\bigr)\sin\omega_{k}(t^{\prime\prime}-t_{0})-\bigl(\mathbf{e}_{{\kappa}}\mathbf{p}_{\mkern-1.5mu t^{\prime}}\bigr)\sin\omega_{k}(t^{\prime}-t_{0})\bigr\}\biggr]}\label{opcomeval}
\end{align}
and 
\begin{align}
 & g^{\prime\prime}(t^{\prime\prime},t^{\prime})=\sum_{\mathbf{k}}\left(\frac{\vert e\vert\eta_{k}}{\hbar\omega_{k}m}\right)^{\!2}\bigl(\mathbf{e}_{{\kappa}}\mathbf{p}_{\mkern-1.5mu t_{0}}\bigr)\nonumber \\
 & \times\biggl[\bigl(\mathbf{e}_{{\kappa}}\mathbf{p}_{\mkern-1.5mu t^{\prime}}\bigr)\sin\omega_{k}(t^{\prime}\!-\!t_{0})\!-\!\bigl(\mathbf{e}_{{\kappa}}\mathbf{p}_{\mkern-1.5mu t^{\prime\prime}}\bigr)\sin\omega_{k}(t^{\prime\prime}\!-\!t_{0})\biggr]\label{comDeval}
\end{align}
The last lines in Eqs. (\ref{opcomeval}) and (\ref{comDeval}) cancel
so that the sum yields 
\begin{align*}
{g(t^{\prime\prime},t^{\prime})=} & {\int_{t^{\prime}}^{t^{\prime\prime}} g^{\prime}(\tau) d\tau+g^{\prime\prime}(t^{\prime\prime},t^{\prime})}
\end{align*}
\begin{align}
g(t^{\prime\prime},t^{\prime}) & =\sum_{\mathbf{k}}\left({\frac{\vert e\vert^{2}}{m\omega_{k}^{2}\varepsilon_{0}V}}\right)\!\frac{1}{2m\hbar}{\int_{t^{\prime}}^{t^{\prime\prime}}}\!\!\bigl(\mathbf{e}_{{\kappa}}\mathbf{p}_{\mkern-1.5mu \tau}\bigr)^{\!2}\!d\tau,\label{geval}
\end{align}
Now we let $V\rightarrow\infty$ by 
\begin{align}
{\sum_{\mathbf{k}}\frac{1}{8\pi^{3}}\!\Delta k^{3}\!=\!{\frac{1}{8\pi^{3}c^{3}}}\int_{0}^{\omega_{r}}\omega_{k}^{2}d\omega_{k}d\Omega_{\mathbf{k}}\mathrm{,}} & \label{vinf}
\end{align}
where $\omega_{r}$ is the cutoff frequency used
for renormalization and $d\Omega_{\mathbf{k}}$ is the angular integration
variable. Clearly the $d\omega_{k}$ part is infinite and $\omega_{r}$
needs to be capped. An estimate can be obtained by following Bethe's
non-relativistic argument for the Lamb shift \citep{bethe1947}; the
electron is non-relativistic, therefore $\omega_{r}\leq mc^{2}/\hbar$
should be capped to frequencies corresponding to non-relativistic
energies. This is intuitive, as the interaction of the electron with relativistic
frequencies would result in relativistic fluctuating motion of the
electron.

We still need to do the angular integral. For that we chose $\mathbf{p}_{t}\parallel k_{z}$
and we chose one polarization state in the plane spanned by $\mathbf{k}$
and $\mathbf{p}_{t}$. Then the second polarization state is automatically
perpendicular to $\mathbf{p}_{t}$ and does not contribute. As the
integrand has only $\theta$-dependence (z-axis angle) we get
\begin{align}
\int d\Omega_{\mathbf{k}}\bigl(\mathbf{e}_{{\kappa}}\mathbf{p}_{{\tau}}\bigr)^{2}=2\pi\mathbf{p}_{\mkern-1.5mu {\tau}}^{2}\int_{0}^{\pi}d\theta\sin^{3}\theta=\frac{8\pi}{3}\mathbf{p}_{\mkern-1.5mu {\tau}}^{2}.\label{angint}
\end{align}
Inserting Eq. (\ref{vinf}) and (\ref{angint}) in (\ref{geval})
yields 
\begin{align}
 & g(t^{\prime\prime},t^{\prime})={\delta_{r}}\,\frac{1}{2m\hbar}{\int_{t^{\prime}}^{t^{\prime\prime}}\!}\!\mathbf{p}_{\mkern-1.5mu \tau}^{2}d\tau\label{gevalint}
\end{align}
\begin{align}
\delta_{r}=\frac{1}{3\pi^{2}}\frac{\vert e\vert^{2}}{\varepsilon_{0}\hbar c}\frac{\hbar\omega_{r}}{mc^{2}}.\label{massre}
\end{align}
Here, $\delta_r$ renormalizes the electron mass by interaction of the continuum electron
with vacuum modes. 

Since the experimentally known mass of the electron
already contains the effects of the electromagnetic field, the phase
term $g(t^{\prime\prime},t^{\prime})$ can effectively be dropped
from our analysis \citep{bethe1947}.

\section{Long electron trajectory return filter \label{sec:Long-electron-trajectory}}

Usually dephasing is accounted for by a term, $\exp \left(- \xi(t-t^{\prime})\right)$, with $\xi(t-t^{\prime})=(t-t^{\prime})/T_{2}$
and $T_{2}$ the dephasing time \citep{coccia2020,kruchinin2019,brown2024}. This increases suppression of HHG
for increasingly higher order electron returns. We use here a different
filter which leaves the first return unfiltered and extinguishes all
higher returns, 
\begin{align}
 & \xi(t-t^{\prime})=0\hspace{2cm}\,\mathrm{for}\,\,t-t^{\prime}\le\frac{T_{0}}{2}\\
 & \xi(t-t^{\prime})=10(t-t^{\prime})/T_{0}\,\,\,\,\,\mathrm{for}\,\,\frac{T_{0}}{2}\le t-t^{\prime}\le T_{0}\\
 & \xi(t-t^{\prime})=\infty\hspace{1.9cm}\mathrm{for}\,\,t-t^{\prime}>T_{0}
\end{align}
with optical period $T_{0}=2\pi/\omega_{0}$ and $\omega_{0}$ the
laser frequency. The filter is confined to one optical cycle, as the
associated convolution operation is numerically expensive.

\section{Differentiation and integration properties of Fourier transforms
\label{sec:Integration-and-differentiation}}

Sec. \ref{sec:Connection-to-interband} relies heavily on the differentiation
and integration properties of Fourier transforms. For the convenience
of the reader we provide them here. 

Throughout this work the convention for the Fourier transform of a
function $G(t)$ in the time domain to the harmonic angular frequency
domain is,
\begin{align}
g(\omega)= & \int_{-\infty}^{\infty}dte^{i\omega t}G(t).
\end{align}
The differentiation and integration properties are
\begin{align}
\int_{-\infty}^{\infty}dte^{i\omega t}\frac{dG(t)}{dt}= & -i\omega g(\omega),\label{eq:diffprop}
\end{align}
\begin{align}
\int_{-\infty}^{\infty}dte^{i\omega t}\int_{-\infty}^{t}dt^{\prime}G(t^{\prime})= & \frac{ig(\omega)}{\omega}+\pi g(0)\delta(\omega),\label{eq:intprop}
\end{align}
respectively \citep{ulaby2016}.

The second term of Eq. (\ref{eq:intprop}) is a DC term that we ignore
throughout this analysis as we account only for harmonics above the
fundamental frequency of the laser beam.